\documentclass[10pt,journal,compsoc]{IEEEtran}
\usepackage{graphicx}
\usepackage{amsmath}
\usepackage[]{xcolor}
\usepackage{tikz}
\usepackage{algpseudocode}
\usepackage{algorithm}
\usepackage{comment}
\usepackage{listings}
\usepackage{multirow}
\usepackage{booktabs}
\usepackage{makecell}
\usepackage{array}
\usepackage{xspace}
\definecolor{codegreen}{rgb}{0,0.6,0}
\definecolor{codegray}{rgb}{0.5,0.5,0.5}
\definecolor{codepurple}{rgb}{0.58,0,0.82}
\definecolor{backcolour}{rgb}{0.95,0.95,0.92}

\lstdefinestyle{mystyle}{
    commentstyle=\color{codegreen},
    keywordstyle=\color{magenta},
    numberstyle=\tiny\color{codegray},
    stringstyle=\color{codepurple},
    basicstyle=\ttfamily\footnotesize,
    breakatwhitespace=false,
    breaklines=true,
    captionpos=b,
    keepspaces=true,
    numbers=left,
    numbersep=5pt,
    showspaces=false,
    showstringspaces=false,
    showtabs=false,
    tabsize=2
}

\algnewcommand\algorithmicforeach{\textbf{for each}}
\algdef{S}[FOR]{ForEach}[1]{\algorithmicforeach\ #1\ \algorithmicdo}

\newcommand{\sysnameaim}[0]{\texttt{AIM}\xspace}
\newcommand{\intTable}{\textit{Interrupt Model Table}}
\let\oldtextbf\textbf
\renewcommand{\textbf}[1]{\oldtextbf{\boldmath #1}}

\newcolumntype{L}[1]{>{\raggedright\let\newline\\\arraybackslash\hspace{0pt}}m{#1}}
\newcolumntype{C}[1]{>{\centering\let\newline\\\arraybackslash\hspace{0pt}}m{#1}}
\newcolumntype{R}[1]{>{\raggedleft\let\newline\\\arraybackslash\hspace{0pt}}m{#1}}

\definecolor{codegreen}{rgb}{0,0.6,0}
\definecolor{codegray}{rgb}{0.5,0.5,0.5}
\definecolor{codepurple}{rgb}{0.58,0,0.82}
\definecolor{backcolour}{rgb}{0.95,0.95,0.92}
 
\lstdefinestyle{mystyle}{
    commentstyle=\color{codegreen},
    keywordstyle=\color{blue},
    numberstyle=\color{codegray},
    stringstyle=\color{codepurple},
    basicstyle=\ttfamily\footnotesize,
    breakatwhitespace=false,         
    breaklines=true,                 
    captionpos=b,                    
    keepspaces=true,                 
    numbers=left,                    
    numbersep=5pt,                  
    showspaces=false,                
    showstringspaces=false,
    showtabs=false,                  
    tabsize=2,
    frame=single,
    xleftmargin=2em,
    framexleftmargin=1.5em,
}

\ifCLASSOPTIONcompsoc
  \usepackage[nocompress]{cite}
\else
  \usepackage{cite}
\fi
\ifCLASSINFOpdf
\else
\fi
\usepackage{url}

\begin{document}
%
\title{AIM: Automatic Interrupt Modeling for Dynamic Firmware Analysis}
%
%
%
%

\author{Bo~Feng,
        Meng~Luo,
        Changming~Liu,
        Long~Lu,
        and~Engin~Kirda
\IEEEcompsocitemizethanks{
\IEEEcompsocthanksitem B. Feng is with School of Cybersecurity and Privacy, College of Computing, Georgia Institute of Technology, Atlanta, GA, 30332, USA.\protect\\
E-mail: bofengwork@gmail.com
\IEEEcompsocthanksitem M. Luo is with School of Cyber Science and Technology, Zhejiang University, Hangzhou, Zhejiang, 310007, China. E-mail: meng.luo@zju.edu.cn
\IEEEcompsocthanksitem C. Liu, L. Lu, and E. Kirda are with Khoury College of Computer Sciences, Northeastern University, Boston, MA, 02115, USA. E-mail: liu.changm@northeastern.edu, l.lu@northeastern.edu, e.kirda@northeastern.edu
}
\thanks{Manuscript received xx, 2022; revised xx, xxxx.\\
This work is partially done when Bo Feng and Meng Luo are with Northeastern University.\\
Corresponding author: Meng Luo.
}
}

%
%

\markboth{Journal of \LaTeX\ Class Files,~Vol.~14, No.~8, August~2015}%
{Shell \MakeLowercase{\textit{et al.}}: Bare Demo of IEEEtran.cls for Computer Society Journals}
%



\IEEEtitleabstractindextext{%
\begin{abstract}

The security of microcontrollers, which drive modern IoT and embedded devices, continues to raise major concerns.
Within a microcontroller (MCU), the firmware is a monolithic piece of software that contains the whole software stack, whereas a variety of peripherals represent the hardware.
As MCU firmware contains vulnerabilities, it is ideal to test firmware with off-the-shelf software testing techniques, such as dynamic symbolic execution and fuzzing. 
Nevertheless, no emulator can emulate the diverse MCU peripherals or execute/test the firmware.
Specifically, the interrupt interface, among all I/O interfaces used by MCU peripherals, is extremely challenging to emulate. 

In this paper, we present \texttt{AIM}---a generic, scalable, and hardware-independent dynamic firmware analysis framework that supports unemulated MCU peripherals by a novel interrupt modeling mechanism. 
\texttt{AIM} effectively and efficiently covers interrupt-dependent code in firmware by a novel, firmware-guided, \textit{Just-in-Time Interrupt Firing} technique.
We implemented our framework in \texttt{angr} and performed dynamic symbolic execution for eight real-world MCU firmware.
According to testing results, our framework covered up to 11.2 times more interrupt-dependent code than state-of-the-art approaches while accomplishing several challenging goals not feasible previously.
Finally, a comparison with a state-of-the-art firmware fuzzer demonstrates dynamic symbolic execution and fuzzing together can achieve better firmware testing coverage. 
\end{abstract}

}

\maketitle

\IEEEdisplaynontitleabstractindextext

%
\IEEEpeerreviewmaketitle

\IEEEraisesectionheading{
\section{Introduction}
\label{sec:intro}
}
\IEEEPARstart{W}{ith}
the proliferation of diverse types of IoT and embedded devices, the security of microcontrollers (or MCU), widely used in these devices, is gaining increasing attention.
A recent security report~\cite{papp2015embedded} has shown that attacks targeting MCU devices may result in not only digital harm but also physical damage.

The MCU is a power-efficient, resource-constrained tiny computer that drives a wide range of modern IoT and embedded devices, such as drones and industrial control systems.
The global MCU market size reached 18.5 billion US dollars in 2021~\cite{mcu_market}. 
Within an MCU, the firmware is pivotal as it refers to a monolithic piece of software that represents the whole software stack of the MCU.
Similar to computer software, MCU firmware contains diverse forms of vulnerabilities as they are also written in memory-unsafe languages, such as C/C++.
The fact that MCU devices are hardly providing powerful defense mechanisms due to inadequate resources makes firmware vulnerabilities more devastating.

To discover MCU firmware vulnerabilities, existing works take advantage of the same software testing techniques (e.g., fuzzing, dynamic symbolic execution) as used in computer software.
They test firmware in an emulator to overcome the tight resource constraint of MCU (which has MHz processor, KB memory, and MB persistent storage). 
A key challenge faced by these mechanisms is that no off-the-shelf emulators can provide comprehensive or highly-automated emulation for the wide range of I/O peripherals equipped on MCU devices, whose functionality ranges from connectivity, sensing, to actuating. 
As a result, MCU firmware cannot execute or be tested in an emulator. 
To enable emulator-based firmware testing, three lines of works have been proposed. 
The first line of works~\cite{zaddach2014avatar,corteggiani2018inception,talebi2018charm,gustafson2019toward} conducts hardware-in-the-loop emulation by forwarding peripheral operations performed in the emulator to a real hardware device.
These approaches are slow and unscalable due to the dependence on real hardware. 
The second line of works~\cite{chen2016towards,clements2020halucinator} gets rid of real hardware by high-level emulation, where they replace the low-level peripheral-dependent code in an MCU firmware with the functionality-equivalent counterpart. 
Nevertheless, they cannot test the code that has been replaced. 
The last line of works~\cite{davidson2013fie,feng2020p2im,cao2020device,mera2021dice,zhou2021automatic,johnson2021jetset,scharnowskifuzzware} handles the unemulated peripherals by novel approaches of modeling peripheral interfaces (i.e., memory-mapped I/O, interrupt, and DMA). 
These works can test the whole MCU firmware without emulating any peripherals or using any peripheral hardware. 
However, these works currently focus on memory-mapped I/O (MMIO) or direct memory access (DMA), and only provide simple support for the interrupt interface (e.g., by firing interrupts at a fixed order and frequency). 
As MCU firmware is event-driven, a significant portion of the firmware that requires complex interrupt sequences cannot be tested.

In this work, we present a novel interrupt modeling mechanism for supporting interrupts while analyzing MCU firmware within emulators.
We build a dynamic firmware analysis framework called \texttt{AIM} that integrates the presented \textit{\textbf{A}utomatic \textbf{I}nterrupt \textbf{M}odeling} mechanism with a popular emulator. 
Using this framework, we can run and test the binary of MCU firmware via dynamic symbolic execution at scale and support the interrupt on demand.
Compared to existing solutions, our method is generic, automated (manual effort is negligible), hardware-independent, and scalable.
It tests firmware without using any real hardware or replacing any firmware components. 
In addition, MCU firmware testing using our framework is efficient (due to just-in-time interrupts) and effective (due to firmware-guided interrupt inference), aiming to maximize code coverage and find more security bugs.

\texttt{AIM} mainly consists of three components: \textit{Interrupt Identification}, \textit{Just-in-Time Interrupt Inference and Firing}, and \textit{Dynamic Symbolic Execution}. 
First, we avoid relying on real hardware by characterizing the effects of diverse types of interrupts on MCU firmware.
Specifically, we propose an MCU-, peripheral-, and OS-agnostic \textit{Interrupt Identification} mechanism, based on our key observations, to accurately and dynamically identify all possible interrupts and their effects on the firmware. 
The interrupt identification process generates an \intTable~as a reference for interrupt inference. 
Second, as our goal is to provide meaningful interrupts for driving MCU firmware to diverse valuable paths, we devise a firmware-guided \textit{Just-in-Time Interrupt Inference and Firing} method, which is capable of accurately recognizing the place that requires the interrupt at run-time and deriving effective interrupt sequences for covering a broad set of unique paths. 
Finally, we extend angr~\cite{shoshitaishvili2016sok}---a dynamic symbolic execution engine for program binaries---to support the operations of diverse peripherals by modeling the peripheral interfaces, especially the interrupt. 
We then integrate interrupt modeling with dynamic symbolic execution to build an effective and efficient dynamic firmware analysis framework.
To the best of our knowledge, we are the first to support the interrupt for dynamic firmware analysis by a fine-grained, on-demand interrupt modeling mechanism.

We evaluated our framework using eight real and full-fledged MCU firmware whose functionalities range from the industrial control system to gateway.
We also implemented two baseline approaches which represent the framework without any interrupt support (\textit{Baseline-No\_INT}) and with the state-of-the-art interrupt support (\textit{Baseline-P\textsuperscript{2}IM}) for dynamic firmware analysis.
We compared our framework with two baseline approaches to demonstrate the effectiveness of our framework for reaching previously uncovered interrupt-dependent code and improving the code coverage of firmware testing.
The result shows that our framework can cover as much as about 45\% of a firmware's basic blocks compared to about 29\% for the state-of-the-art approach (\textit{Baseline-P\textsuperscript{2}IM}).
The improvements achieved by our framework are also significant in that it can cover up to 11.2 times more interrupt-dependent code than \textit{Baseline-P\textsuperscript{2}IM} while fulfilling a few impressive features, such as generic, fully-automated, hardware-independent, and scalable.
In addition, we demonstrate dynamic symbolic execution and fuzzing together can potentially achieve better firmware testing coverage by comparison with a state-of-the-art firmware fuzzer (Fuzzware~\cite{scharnowskifuzzware}). 
We open source our tool and dataset at \url{https://github.com/bofeng17/AIM-Interrupt-Modeling}. 
\section{Overview}
\subsection{Background}

\begin{figure}[!ht]
    \centering
    \includegraphics[width=0.8\columnwidth]{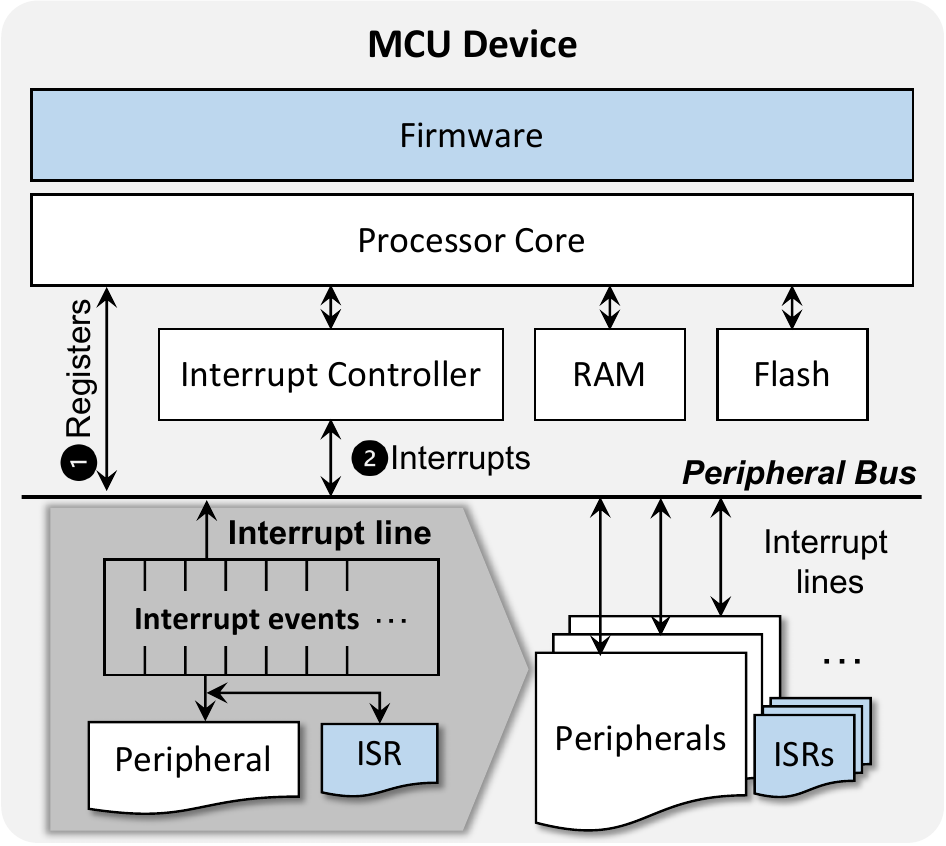}
    \caption{Architecture of MCU devices with MMIO and complex interrupt I/O interface (software components are marked with blue color; all the other components are hardware components)}
    \label{fig:architecture}
\end{figure}

MCUs are used across a broad range of applications. 
In this section, we give a brief introduction to the architecture of MCU devices and their peripheral I/O interfaces.
As shown in Figure~\ref{fig:architecture}, MCU devices can be generally divided by software and hardware.
The firmware, which represents the software part, includes the whole software stack of MCU devices, including an optional real-time OS (RTOS), drivers, libraries, and application logic.
As for the hardware part, it mainly consists of the processor core, interrupt controller, RAM, flash, and a set of peripherals.
To perform I/O with a wide range of peripherals, memory-mapped registers and interrupts are two primary approaches.
Through memory-mapped registers, or MMIO, the processor core can configure, check the status of, and exchange data with peripherals. 

The processor core can also obtain peripheral states through interrupts.
Specifically, an MCU device can contain a number of interrupt lines (hundreds on ARM Cortex-M MCUs), each of which is assigned to a peripheral. 
When peripheral states change, such as time elapsed or data reception, the peripheral will notify the processor through its interrupt line by firing interrupts. 
All interrupt lines are managed by a special peripheral called interrupt controller on behalf of the processor.
Interrupt controller is responsible to schedule all received interrupts and determine when to interrupt the processor based on the priority of different tasks. 
To serve an interrupt, the processor looks up the interrupt vector table and switches the application to the interrupt service routine (ISR) associated with the interrupt line that has transmitted peripheral-state changes.
Afterward, ISR, which belongs to the firmware, handles the interrupt by accessing some memory-mapped peripheral registers.
Finally, the processor resumes the normal execution of applications.

It is worth noting that each interrupt line can multiplex a variety of peripheral events, such as data transmission, data receiving, and error reporting.
The ISR for an interrupt line can contain different event handlers for diverse interrupt events.
As a result, it is the ISR that identifies which event has triggered an interrupt via the memory-mapped peripheral status register, and then invokes the corresponding event handler.

\subsection{Motivation}
For dynamic firmware analysis, it is indispensable to support diverse peripheral operations to trigger specific pieces of code within the firmware.
That being said, at least peripheral I/O interfaces through which peripheral is handled entail to be supported, if not the entire peripheral.
Otherwise, a firmware may fail to trigger certain application logic, cease to boot, or even crash, causing dynamic firmware analysis ineffective.
For frequently utilized peripheral I/O interfaces, namely memory-mapped registers and interrupts, the interrupt is more complex and nondeterministic, thereby hampering its support in emulators.

\lstset{style=mystyle}
\begin{lstlisting}[language=C,caption={Code snippet of MCU firmware during booting. }, label=listing:delay, escapechar=|]
volatile uint32_t uwTick;

int boot(){
  ...
  configure_peripheral();
  delay(5); |\label{ld:delay1}|// wait for peripheral hardware
  ... 
}

void delay(uint32_t ms){ |\label{ld:delay_s}|
  uint32_t start = uwTick; // uwTick is modified by SysTick interrupt |\label{ld:check1}|
  while (uwTick - start < ms) {} // exit after ms milliseconds  |\label{ld:check2}|
} 
\end{lstlisting}

In Listing~\ref{listing:delay}, we show the code snippet of an MCU firmware to demonstrate the necessity of firing interrupts during dynamic firmware analysis. 
This example, representing that a firmware could not boot without interrupts, is quite common among a plethora of MCU devices. 
During booting, the firmware configures peripherals and then invokes \texttt{delay()} function (line~\ref{ld:delay1}) to wait for peripheral operations since peripherals are usually slower than the processor by several orders of magnitude. 
The \texttt{delay()} function keeps spinning until a specified amount of time has elapsed (line~\ref{ld:delay_s}).
Specifically, it relies on the timer interrupt to be fired and to increase the value of \texttt{uwTick} global variable (line~\ref{ld:check2}).
The firmware will get stuck in the loop until the timer interrupt has been fired by several times.  
Without interrupts, a firmware is unable to boot successfully and thus may not be possible to be analyzed. 
Therefore, it is inevitable to present an approach for firing interrupts effectively to enable in-depth dynamic firmware analysis within emulators.

Since there lacks an emulator that is capable of comprehensively supporting peripheral operations, dynamic analysis techniques are hard to be applied to identify vulnerabilities in MCU firmware. 
To support peripheral operations, prior work has mainly focused on memory-mapped registers and DMA, leaving the interrupts poorly supported.
The first line of related works is hardware-dependent and relies on real hardware to support peripheral operations.
Specifically, Avatar~\cite{zaddach2014avatar}, Inception~\cite{corteggiani2018inception}, and Charm~\cite{talebi2018charm} trigger interrupts in real devices and then forward them to the counterparts of tested firmware hosted in an emulator.
However, forwarding interrupts from real devices can be very slow.
To overcome this problem, Pretender~\cite{gustafson2019toward} is presented to generate interrupts directly inside an emulator after obtaining interrupt models from real MCU devices.
A common limitation with all of the above approaches is that it entails significant manual efforts to bridge the gap between an emulator and a real MCU device. 

The second line of work tries to infer possible interrupts without relying on real hardware.
For example, FIE~\cite{davidson2013fie} tries to exhaustively trigger all possible interrupts after executing each instruction, but it causes the symbolic execution engine to quickly encounter the state explosion issue. 
P$^2$IM~\cite{feng2020p2im}, Laelaps~\cite{cao2020device}, $\mu EMU$\cite{zhou2021automatic}, Jetset\cite{johnson2021jetset}, and Fuzzware\cite{scharnowskifuzzware} utilize naive interrupt modeling, which fires interrupts with a fixed order and frequency. 
The problem is that it is difficult to cover a significant portion of firmware code, such as those that require complicated interrupt sequences. 
Firmadyne~\cite{chen2016towards} and HALucinator~\cite{clements2020halucinator} get rid of modeling interrupts by adopting high-level emulation. 
They replace the peripheral-dependent code in a firmware with a peripheral-independent, functionally equivalent counterpart. 
Nonetheless, all of these approaches suffer from low code coverage. 
Finally, the limitations of prior work motivate us to present an automatic, fine-grained, and hardware-independent interrupt modeling for effectively supporting intricate interrupts.

\subsection{Open Challenges}
With a plethora of MCU devices and their firmware to analyze, it is non-trivial to support automated interrupt firing in emulators across a wide range of peripherals and MCU implementations by not relying on any source code or real hardware of MCU devices. 
Specifically, we encounter the following challenges:

\vspace{0.2em}
\noindent\textbf{Hardware Dependence:} 
MCU is extremely resource-constrained with MHz processor, KB memory, and MB persistent storage. 
Therefore, it is impractical, or even impossible, to test MCU firmware on real devices. 
Besides, when real hardware is used, MCU firmware analysis suffers from poor scalability since it is infeasible to add extra analysis instances without purchasing new hardware (as demonstrated in~\cite{zaddach2014avatar,corteggiani2018inception,talebi2018charm,gustafson2019toward}). 
In consequence, it is necessary to perform hardware-independent firmware testing (i.e., not using any real MCU hardware) in an emulator. 
However, it is a very challenging task to emulate the diverse peripherals equipped on MCUs (e.g., sensors and actuators) during dynamic firmware analysis.
To solve this issue, we propose an Automatic Interrupt Modeling mechanism to generate interrupts in lief of the unemulated peripherals during firmware analysis. 
Using our mechanism, firmware code that depends on interrupts can be executed and analyzed in emulators that do not emulate any peripherals. 

\vspace{0.2em}
\noindent\textbf{Diversity of MCUs, Peripherals, and OSes:}
Unlike desktop and mobile platforms, both the hardware and software of MCU devices exhibit great diversity. 
There are several major MCU vendors in the market, each of which has produced hundreds of MCU models.
In addition, each MCU device is provided with a variety of peripheral types (e.g., sensors, actuators, buses, connectivity, and storage) for interacting with the cyberspace and the physical world. 
Examples of MCU peripherals include \texttt{USART}, \texttt{I2C}, \texttt{SPI}, \texttt{ADC}, etc.
Although all these peripherals are capable of firing interrupts, they leverage interrupts in different ways for tackling distinct events. 
For example, \texttt{TIMER} reports time elapse whereas \texttt{USART} informs data reception. 
Due to the diversity of MCU models, an MCU device could instantiate peripherals, as well as utilize interrupts, differently. 

Similar to hardware, MCU software varies a lot from vendor to vendor, or even device to device, as there is no single OS dominates the market. 
MCU firmware may be built on any OS or even bare metal.
To name a few, FreeRTOS~\cite{FreeRTOS}, VxWorks~\cite{VxWorks}, ARM Mbed~\cite{MBedOS}, Zephyr~\cite{Zephyr}, NuttX~\cite{NuttXOS}, and ChibiOS~\cite{ChibiOS} are all popular OSes for MCU devices.
Due to the diversity and discrepancies among MCU models, peripherals, and OSes, it is, therefore, necessary to incorporate an automatic and generic (i.e., MCU, peripheral, and OS-agnostic) approach for supporting interrupts in emulators.

\vspace{0.2em}
\noindent\textbf{Complexity of Interrupts:} 
To improve code coverage during dynamic firmware analysis, it is important to cover interrupt-dependent paths in an MCU firmware, which can only be triggered when specific interrupt sequences are fired. 
However, generating appropriate interrupt sequences, when necessary, is non-trivial.
Interrupt sequences are a group of interrupts, each of which is annotated with the timing of interrupt firing and a proper input for the interrupt.
Due to the nondeterminism of interrupt firing timing and the complexity of interrupt sequences, it is infeasible to exhaustively generate all possible interrupt sequences and their firing timings.
Therefore, the challenge here is timely and precisely determining when to fire an interrupt and what interrupt to fire on behalf of a peripheral, and then efficiently and effectively generating interrupt sequences for dynamic firmware analysis.

Moreover, determining only the interrupt is not enough.
Each MCU device can contain dozens of interrupt lines, each of which is associated with a peripheral. 
Due to the limited number of interrupt lines, MCU peripherals usually multiplex multiple events (e.g., data reception, data transmission, error reporting, etc.) on the same interrupt line. 
Both interrupt lines and events can affect MCU firmware execution and analysis in different ways. 
As such, another practical challenge of interrupt firing is from event multiplexing.
Since events are independent from each other and multiple events can occur simultaneously, it is, therefore, necessary to fulfill fine-grained interrupt firing (at event granularity) and determines not only the interrupt line, but also the specific event on that interrupt line.

\vspace{0.2em}
\noindent\textbf{Automation:}
Achieving automation while supporting interrupts is important due to the following reasons.
First, interrupt firing is time-sensitive during dynamic firmware analysis. 
Second, it is infeasible to manually support all kinds of interrupts due to the diversity of MCU models, peripherals, and OSes.
Finally, the complexity of interrupts has made supporting interrupts an error-prone task, if manual efforts are provided. 
As a result, a significant challenge is automatically supporting interrupts for performing dynamic firmware analysis. 

\vspace{0.2em}
\noindent\textbf{Unavailability of Source Code:} 
The source code of MCU firmware is unavailable most of the time.
Therefore, a practical challenge is supporting interrupts during dynamic firmware analysis without relying on any source code of MCU firmware.

\subsection{Approach Overview}
Our goal is to build a generic, fully-automated, and hardware-independent interrupt modeling system for supporting peripheral operations in emulators, which is capable of getting rid of real hardware and overcoming the uncertainty and diversity across different MCU models, peripherals, and OSes, as well as the complexity of interrupts.
Especially, we present a novel, just-in-time (JIT) interrupt firing mechanism that can significantly improve code coverage of dynamic firmware analysis by effectively and efficiently covering the interrupt-dependent paths of MCU firmware.

\begin{figure}[!ht]
    \centering
    \includegraphics[width=0.9\columnwidth]{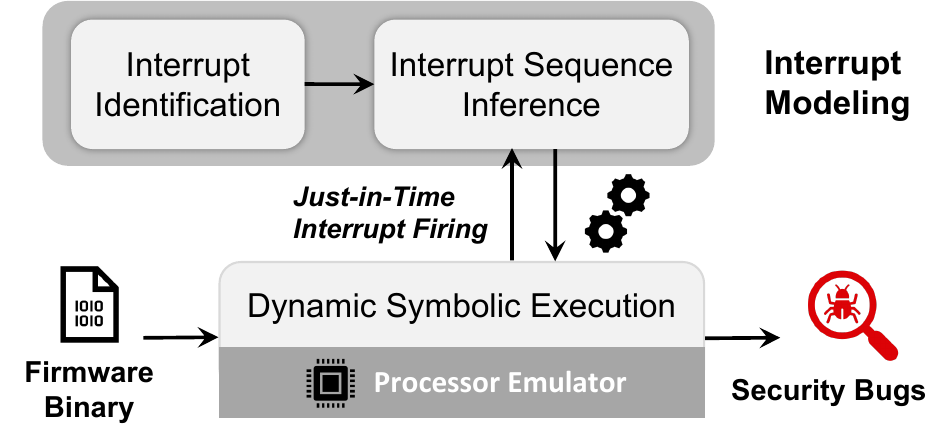}
    \caption{Framework overview}
    \label{fig:approach}
\end{figure}

Our approach is implemented in a framework called \sysnameaim~to support dynamic analysis of MCU firmware in emulators while fulfilling automatic, just-in-time interrupt firing.
As illustrated in Figure~\ref{fig:approach}, \sysnameaim~mainly consists of two components: 1) an automated interrupt modeling system, and 2) a dynamic symbolic execution system built based on emulators.

Given an MCU firmware binary, the dynamic symbolic execution system will conduct normal symbolic execution aiming to identify security bugs within the firmware.
In the meanwhile, the interrupt modeling system starts creating an \intTable~via interrupt identification.
This table describes what interrupts are being used by the firmware and how different interrupts can take effects on the states of a firmware.  
We later utilize this table to infer possible interrupt sequences at run-time according to the execution of the firmware.
Specifically, our system first needs to automatically identify active interrupt lines and the enabled events multiplexed on each interrupt line at the firmware initialization stage and during the run-time.
By analyzing the \textit{Interrupt Service Routine} associated with each interrupt line, our system could further determine the unique effects that diverse interrupts can cause on the states of a firmware, such as the modification patterns to particular global variables.
We will present the details of interrupt identification in \S\ref{sec:identify_int}. 

Immediately when the firmware execution entails interrupts (i.e., reading data from some global variables), the interrupt modeling system can automatically and effectively generate appropriate interrupt sequences and fire them right before a firmware reads data from global variables, thereby achieving just-in-time interrupt firing.
To infer interrupt sequences, it is not enough to merely consider the interrupt line.
Instead, we are able to infer fine-grained, event-level interrupts so that interrupt firing can be very precise and cover interrupt-dependent paths that require complicated interrupts.
In addition, to maximize code coverage and be efficient while generating interrupt sequences, we perform firmware-guided interrupt sequence inference by analyzing the usage of required interrupts and prioritizing interrupt sequences that can lead a firmware to as diverse paths as possible. 
As a result, our fine-grained, context-aware interrupt sequence inference and firing enable the dynamic symbolic execution system to cover extremely complicated interrupts, as needed by a firmware, and accomplish high code coverage during dynamic firmware analysis.
We will elaborate on how we infer interrupt sequences and fire interrupts just-in-time in \S\ref{sec:int_firing}. 

Finally, the dynamic symbolic execution system provides the environment for conducting symbolic execution on MCU firmware by extending angr~\cite{shoshitaishvili2016sok}, a binary analysis framework that allows symbolic execution, to support peripheral interfaces, namely registers and interrupts.
While driving dynamic symbolic execution for a firmware, our system also needs to tackle the path explosion problem.
To mitigate this problem, we apply both path filtering and scheduling methods to maximize code coverage, thereby benefiting the identification of security bugs.
We will discuss the details of dynamic symbolic execution in \S\ref{sec:dse}. 

\section{Interrupt Modeling}
\label{sec:design}
To perform dynamic firmware analysis in an emulator that does not support peripherals or their interrupts, we design an automated interrupt modeling mechanism which triggers interrupts just-in-time when one or more interrupts are needed.
The interrupt modeling mechanism enables dynamic symbolic execution to effectively cover diverse interrupt-dependent code in the firmware. 
In this section, we will present the two main components of interrupt modeling, namely interrupt identification and interrupt sequence inference and firing.

\subsection{Interrupt Identification}
\label{sec:identify_int}
\begin{figure}[!t]
    \centering
    \includegraphics[width=\columnwidth]{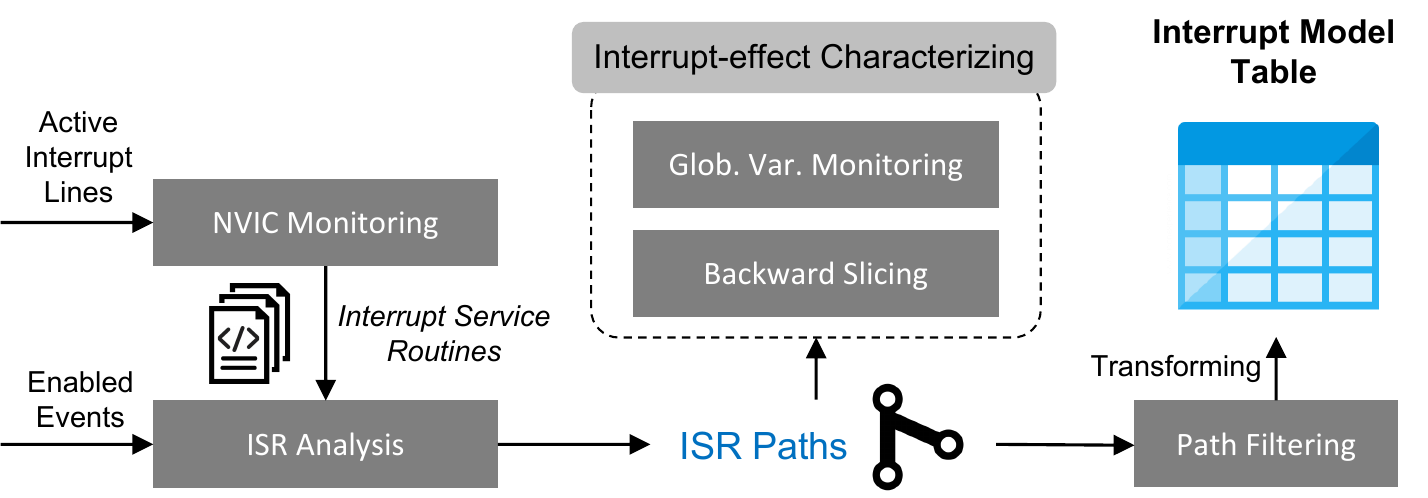}
    \caption{Interrupt identification workflow}    
    \label{fig:int_identification}
\end{figure}

Interrupt identification aims to automatically and dynamically identify the interrupts that are being used by the firmware at runtime and characterize their effects. 
It first performs fined-grained identification at interrupt line and event granularity. 
Then, it characterizes the interrupt effects, which, if fired, can drive the firmware to different pieces of code, by understanding what and how particular global variables are modified by the interrupt. 
Finally, it generates and maintains a dynamically-updated \intTable~to describe all types of currently enabled interrupts/events and their effects, which is later used in interrupt firing. 
Figure~\ref{fig:int_identification} shows the workflow of interrupt identification. 

\subsubsection{Fine-grained Interrupt Line \& Event Identification}
We perform fine-grained interrupt identification at the interrupt line and event level because, as shown in Figure~\ref{fig:architecture}, several events, such as data reception and error occurred, can multiplex on an interrupt line. 
When firing an interrupt, we need to decide not only the interrupt line but also the specific events that trigger the interrupt.

Both interrupt lines and their events are flexibly enabled/disabled at run-time and they are by default disabled to avoid unexpected behaviors and save MCU power. 
Mistakenly firing interrupts through inactive interrupt lines or disabled events usually has no effect.
But in some cases, it can crash the firmware (when the firmware does not check if an event is enabled before handling it through calling the corresponding event handler via a function pointer which is a null pointer because the event is not enabled or configured). 
Therefore, we identify the enabled interrupt lines and events at runtime and assure that we only fire them at the interrupt firing phase.

\textbf{Interrupt Line Identification.} 
We identify active interrupt lines by monitoring the interrupt controller, i.e., \textit{Nested Vectored Interrupt Controller (NVIC)} on ARM Cortex-M MCUs.
Specifically, we observe that MCU firmware enables (or disables) interrupt lines by writing to NVIC's \texttt{NVIC\_ISERx} (or \texttt{NVIC\_ICERx}) register, each bit of which represents an interrupt line. 
In addition, the NVIC is included in the standard ARM Cortex-M architecture~\cite{cortexm3-nvic}, and thus it allows us to identify active interrupt lines in an MCU-agnostic way.

\textbf{ISR Analysis-based Event Identification.}
However, it is more challenging to identify what events are enabled for a given interrupt line because events are managed by individual peripherals which are not emulated. 
We realize the ISR, which is responsible for handling interrupts triggered by diverse events, encodes the event configuration and handling information. 
Therefore, we design a novel algorithm to extract event information from an ISR. 

\lstset{style=mystyle}
\begin{lstlisting}[language=C,caption={Code snippet of a real-world USART ISR implemen-\\tation on STM32F103 MCU (truncated for brevity). }, label=lst:running_eg, escapechar=\\]
void HAL_UART_IRQHandler(UART_HandleTypeDef *huart){
  uint32_t isrflags   = READ_REG(huart->Instance->SR);
  uint32_t cr1its     = READ_REG(huart->Instance->CR1);
  uint32_t cr3its     = READ_REG(huart->Instance->CR3);

  // If data has been received
  if(((isrflags & USART_SR_RXNE) != RESET) && ((cr1its & USART_CR1_RXNEIE) != RESET)) {
    UART_Receive_IT(huart);
    return;
  }

  // If data has been transmitted
  if(((isrflags & USART_SR_TXE) != RESET) && ((cr1its & USART_CR1_TXEIE) != RESET)) {
    UART_Transmit_IT(huart);
    return;
  }

  // If some errors occur 
  uint32_t errorflags = isrflags & (USART_SR_PE 
| USART_SR_FE | USART_SR_ORE | USART_SR_NE);
  if((errorflags != RESET) && (((cr3its & USART_CR3_EIE) != RESET) || ((cr1its & USART_CR1_PEIE) != RESET))) {
    ... // error handling
    return;
  }
}
\end{lstlisting}

Listing~\ref{lst:running_eg} presents the code snippet of a real-world USART ISR implementation. 
We use it as a running example to explain our algorithm.
An ISR generally follows the ``\textbf{check if an event is triggering -$>$ check if an event is enabled -$>$ handle the event}" paradigm to handle an event. 
Take the data receiving event as an example, first, the ISR checks if this event is triggering the interrupt being handled by evaluating \texttt{(isrflags \& USART\_SR\_RXNE) != RESET} at line 7.
Specifically, it checks if the \texttt{USART\_SR\_RXNE} flag in \texttt{SR} (a Status Register) is set or not. 
If it is set, the ISR believes the data-receiving event triggered this interrupt. 
Then, the ISR checks if the data receiving event is enabled (i.e., allowed to trigger interrupts) by evaluating \texttt{(cr1its \& USART\_CR1\_RXNEIE) != RESET} at line 7. 
Specifically, it checks the \texttt{USART\_CR1\_RXNEIE} flag in \texttt{CR1} (a Control Register) is set or not. 
If it is set, the ISR believes the data-receiving event is enabled and allowed to trigger interrupts. 
Finally, the ISR will handle the event by calling \texttt{UART\_Transmit\_IT} at line 8. 
In the same way, the ISR handles the data-transmission event (line 13-16) and error events (line 19-24). 
In summary, there are 6 events in total: \texttt{RXNE}, \texttt{TXE}, \texttt{PE}, \texttt{FE}, \texttt{ORE}, \texttt{NE}, while the last 4 are error events. 
These events map to bit 5, bit 7, bit 0, bit 1, bit 3, and bit 2 of \texttt{SR} respectively. 
However, there are only 4 switches in \texttt{CR} controlling the event enable/disable: \texttt{RXNEIE}, \texttt{TXEIE}, \texttt{PEIE}, \texttt{EIE}. 
The last switch controls the last three error events. 
These switches map to bit 5 of \texttt{CR1}, bit 7 of \texttt{CR1}, bit 8 of \texttt{CR1}, and bit 0 of \texttt{CR3} respectively.

We summarize the key observations as follows. 
First, an MCU peripheral stores the event configuration (i.e., enable/disable) in its control registers (\texttt{CR}). And the peripheral indicates which events are triggering an interrupt by flags in the status register (\texttt{SR}).
Second, the ISR, which is associated with an interrupt line, checks \texttt{CR} and \texttt{SR} to see if the event is enabled and if the event is triggering the current interrupt before handling an event\footnote{We rely on existing register categorization mechanism (e.g., P\textsuperscript{2}IM~\cite{feng2020p2im}) to categorize registers. For \texttt{C\&SR}. which has a mix of \texttt{CR} and \texttt{SR} bits, the categorization mechanism decides, for each bit, whether it belongs to \texttt{CR} or \texttt{SR} for us.}.

Based on this information, we design a dynamic symbolic execution-based ISR analysis algorithm to identify events. 
The algorithm starts from the ISR entry point and terminates after all paths return from the ISR function.
During dynamic symbolic execution, it preserves the value of \texttt{CR} registers that is previously written by the firmware to record the event configurations. 
But it symbolizes the \texttt{SR} value to mimic that different events are triggering interrupts because a symbolic value can be any of the candidate values. 
During dynamic symbolic execution, paths will fork at \texttt{SR} flag checks.
Using data receiving event as an example, the check \texttt{(isrflags \& USART\_SR\_RXNE) != RESET} at line 7 forks two paths: one indicates the event is triggering the current interrupt, the other is not. 
For the former path, if the event has been enabled, the corresponding event handler will be invoked at line 8, where the interrupt causes effects on the firmware by changing its states. 
We collect and analyze the effects in the next section. 

In this way, we get a collection of symbolic execution paths. 
Each path corresponds to a specific event combination, e.g. (\texttt{RXNE}, $\neg$\texttt{TXE}, $\neg$\texttt{PE}, $\neg$\texttt{FE}, $\neg$\texttt{ORE}, $\neg$\texttt{NE}).  
Note that multiple events can happen simultaneously in a single interrupt. 
We solve the symbolic value of \texttt{SR} by Z3 constraint solver to get the concrete value containing event flags.
By ISR analysis, we can recover all the events that are enabled, e.g., data receiving, data transmission, or error reporting. 
For events that are not enabled, we may miss it if the ISR checks if the event is enabled before checking if the event is triggering because symbolic execution won't fork paths for such events due to the short-circuited if condition evaluation.

\textbf{ISR Analysis Triggers.}
We launch ISR analysis when a new interrupt line or event becomes active. 
The NVIC monitoring informs us of new active interrupt lines. 
For events, we monitor the value of peripheral \texttt{CR} registers,  which manage event enable/disable.

As peripherals are not emulated, we are unaware of which peripheral and its \texttt{CR} an ISR is associated with. 
We identify such associations at the first ISR analysis by identifying the most frequently accessed peripheral by the ISR.

In the next section, we will leverage these ISR paths to further analyze the effects of interrupts and events on the MCU firmware, which is essential for determining what interrupts to trigger during firmware testing.

\subsubsection{Understanding Interrupt Effects}
The interrupt, which is necessary for the execution of MCU firmware, can be triggered by the enabled events of an active interrupt line and then handled by a unique ISR. 
Therefore, the natural questions are: 1) what are the effects of a triggered interrupt on an MCU firmware? and 2) how can these effects help our interrupt inference? 

To answer these questions, we compare MCU firmware's states before and after executing ISRs and observe that ISRs have modified a few global variables in the memory (i.e., RAM).
In addition, after exiting the ISR, it is these global variables that drive an MCU firmware towards different paths.
Therefore, given an interrupt line, it is important to understand how the above global variables are related to the interrupts that are triggered by diverse events.
Specifically, we have to first identify what global variables are modified by each of the ISR event handling paths and then how these global variables are modified, namely the modification pattern. 
As such, our goal here is to generate an \intTable~to describe the relationship between interrupts and global variables. 
Later, we may refer to the \intTable~to infer what interrupts to support for dynamic firmware analysis.

First, we figure out the memory region that stores all global variables so that we can identify the global variables modified by the ISR via monitoring this region. 
To this end, we leverage reverse engineering techniques to analyze an MCU firmware's \texttt{Reset\_Handler} function, which is a standard function included in the ARM Cortex-M architecture~\cite{cortexm3-nvic}. 
The function performs various initialization for the firmware, including all global variables. 
As long as we have figured out the memory region for global variables, we can monitor the modifications towards that memory space while analyzing the diverse ISR paths, in order to identify the address of modified global variables.

Second, to understand how global variables are modified, we leverage dynamic backward slicing technique to analyze the event handling code within diverse ISR paths obtained earlier.
In the event handling code of an ISR, the values written to global variables are determined by multiple instructions. 
Therefore, we slice the program for each ISR path backward, starting from the instruction that finally writes to a global variable, to obtain all instructions that affect the values of that global variable.
From those instructions, we generate a formula, which composes the value of a global variable.
To be able to infer interrupts from the above formula, we also extract the modification pattern of global variables.
We repeat this process for all global variables modified within an ISR path.

To explain why extracting the modification pattern of global variables can help interrupt inference, we will show a concrete example.
When we conduct dynamic symbolic execution on an MCU firmware, it entails the value of a global variable to be four in order to enter a path.
According to our analysis of the ISR, this global variable is modified by adding one to its current value. 
If we hope the global variable, which equals one, changes to four, we need to invoke the ISR three times to satisfy the requirement.
That is to say, we should provide a sequence of three same interrupts.
The example reminds us that, depending on the modification pattern, we may infer different interrupt sequences to make a global variable reach the same value.
As such, we define four modification patterns for global variables as the following: (1) \textit{constant assignment:} the write instruction assigns a constant value to a global variable;
(2) \textit{self-referral:} the write instruction reads a value from a global variable and then updates the global variable itself (e.g., $val = val + 1$);
(3) \textit{data reception:} the write instruction reads data from a peripheral's data register (\texttt{DR}) and stores it into a global variable;
and (4) \textit{other cases}.
This category covers any other complicated cases.
One example is that the write instruction consumes data from multiple global sources and writes to another global variable.

\subsubsection{Path Filtering}
Among diverse ISR paths, the event handling code may modify different sets of global variables or make modifications towards the same global variables differently.
However, it turns out that there are still some ISR paths that make identical modifications, meaning that they modify the same set of global variables with identical formulas.  
As ISR paths resulting in the exactly same values for global variables will lead an MCU firmware to a common execution path during dynamic symbolic execution, it is meaningless to test all of the duplicated ISR paths.
Hence, for each unique set of global variables, we determine to keep only one ISR path when multiple paths make identical changes over these global variables.
Besides that, we also consider a special case where an ISR path, which handles multiple events sequentially, makes the same modifications as another set of ISR paths collectively do.
For example, the first and the second ISR paths modify four global variables for \texttt{event\_1} and six global variables for \texttt{event\_2} respectively, while the third ISR path modifies the same ten global variables.
In this case, our policy is to filter out the third ISR path.
The reason is that even though both the first two ISR paths combined and the third ISR paths result in identical changes, the first two will cause fewer side effects when we fire interrupts that are handled by that ISR path during dynamic firmware analysis.
We will explain more details about side effects in Section~\ref{sec:int_firing}.
As a result, by applying our path filtering strategy, the performance of dynamic symbolic execution is improved significantly as we can drive MCU firmware to previously uncovered paths more effectively.

\subsubsection{Interrupt Model Table}
Up to now, we have obtained the relationships between interrupt lines, events, and global variables by analyzing distinct ISRs.
The problem here is that only one global variable is later read and decides the execution of MCU firmware.
As a consequence, we transform what we have learned to an \intTable~so that our framework can efficiently search in the table to infer the interrupts that it needs to provide during dynamic firmware analysis.
Specifically, \intTable~allows us to search candidate ISR paths based on a global variable.
For each unique global variable, it follows with a list of ISR paths, each of which is associated with an \textit{interrupt line}, \textit{\texttt{SR} value}, \textit{formula}, \textit{modification pattern}, and \textit{side effects}. 
The \texttt{SR} value indicates the requirement (i.e., events) of entering an ISR path.
The formula of the write instruction towards global variables and the modification pattern helps us determine the sequence of interrupts.
The side effects tell us what other global variables are also modified along with the target.
When a global variable is needed while conducting dynamic symbolic execution on MCU firmware, we randomly choose one ISR path from those trigger the least side effects.

\subsection{Just-in-Time Interrupt Sequence Inference and Interrupt Firing}
\label{sec:int_firing}

\begin{figure}[!t]
    \centering
    \includegraphics[width=\columnwidth]{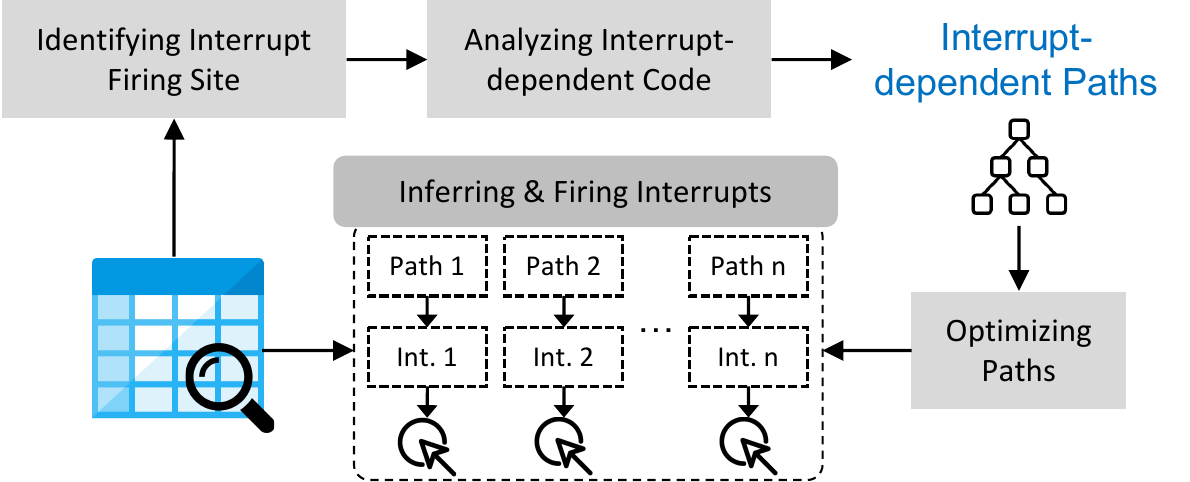}
    \caption{Workflow of just-in-time interrupt sequence inference and firing}
    \label{fig:interrupt_inference}
\end{figure}

To analyze MCU firmware with dynamic symbolic execution, we need to ensure that MCU firmware can execute smoothly and dynamic firmware testing can be effective, by feeding interrupts using our framework.
To this end, we have made two goals, namely achieving accurate and just-in-time interrupts.

The execution of an MCU firmware normally requires interrupts, which can modify the states of the firmware through particular global variables and lead it to go along different paths. 
However, an interrupt may be fired at any time and only the interrupt that is fired at the right time is meaningful.
It is, therefore, useless to trigger interrupts that will cause an MCU firmware to follow the paths that have already been covered.
Hence, achieving just-in-time interrupt firing, which means we only fire the interrupt when an MCU firmware is calling for it, can bring many benefits for us.
Next, to be able to effectively cover untested paths, it is reasonable to consider the precondition of entering distinct paths.
In other words, we have to figure out what value range for the global variable will result in what path.
Furthermore, both the current state and the effects of interrupts can contribute to the value of global variables.
In order to reach as many paths as possible, our interrupt sequence inference is guided by the requirements of different paths.
Specifically, to infer what interrupts are needed, we have to identify the gap between the current state and the precondition of entering different paths.
This process should be fully-automated and efficient.
In a word, in this work, we make the design goals fundamentally different from the existing works.
That is to say, our interrupt inference and firing is firmware-centric and efficiency-driven, unlike the existing approaches that do not take into account the internals of MCU firmware.

As shown in Figure~\ref{fig:interrupt_inference}, the dynamic support of interrupts mainly involves four steps. 
Before determining what interrupts to support, we have to first identify the interrupt firing site, namely which code relies on interrupts, while MCU firmware is executing.
For this purpose, we monitor whether the MCU firmware is reading any of the global variables that appear in the \intTable~(constructed in \S\ref{sec:identify_int}). 
Furthermore, we are also able to obtain how interrupt modifications can affect the global variable.
As mentioned earlier, when the interrupt is handled by ISR, it will modify multiple global variables. 
The modifications towards the global variables other than what we target are likely to cause unnecessary side effects.
As different interrupt modifications may involve a different group of global variables, including the one we target, we have to prioritize and determine the one that triggers the least side effects.
This means that the selected one should modify the least number of global variables, thereby causing the minimum side effects. 
The good news is that side effects rarely happen since the event handling code within ISR normally changes a dedicated set of global variables.

Second, once we find an interrupt firing site, we analyze the MCU firmware code to identify a set of unique paths that require distinct interrupts.
Specifically, we leverage the symbolic execution to analyze the following code of the interrupt firing site.
To this end, we symbolize the global variable and start symbolic execution from the interrupt firing site. 
During the symbolic execution, an MCU firmware may execute along different paths depending on the value of the symbolized global variable.
We terminate the symbolic execution until all paths have executed 30 basic blocks.

However, as we checked the newly discovered basic blocks, we noticed that some symbolic execution paths happen to be covered by other paths.
In other words, among all the paths, the newly discovered basic blocks of one path may be included in other paths.
In this case, we favor the path that contains the larger number of previously uncovered basic blocks and eliminate other paths.
But if a symbolic execution path covers at least one basic block that is not covered by others, we will keep that path.
By conducting path optimization, we avoid wasting time on ``useless" paths by cutting the number of paths that entail inferring interrupts, thereby improving the efficiency of supporting interrupts.
For all the remaining paths, we collect the constraints for the symbolized global variable, which indicate the pre-condition of entering each of these paths. 

The last step is to infer the interrupt sequence for each symbolic execution path.
As discussed before, we maintain some metadata (e.g., modification pattern) in the \intTable~for each global variable in order to demonstrate how an interrupt can modify a global variable.
Here, we leverage this information to infer a sequence of concrete interrupts that can change the value of the global variable to what satisfies the requirement of a symbolic execution path. 
In the simplest case, the current value of the global variable happens to fit the requirement of a path, meaning that there is no need to fire an interrupt.
In other cases, it is necessary to fire one or more interrupts to trigger modifications toward global variables. 
To infer an interrupt sequence, we utilize different strategies based on the modification pattern associated with a global variable. 
For global variables in constant assignment or data reception categories, we can simply generate one interrupt to fire and observe if that will lead us to one of the symbolic execution paths.
For global variables in self-referral or other categories, we adopt a brute-force method where we try to increase the number of interrupts by one every time and check the value of the global variable against the constraint solver to gauge if that meets the constraints.
This process is repeated either until we reach the maximum times of trial or we have generated an interrupt sequence for the symbolic execution path.
As a result, we would be able to know the length of the interrupt sequence for the path if we managed to infer an interrupt sequence for it.
Finally, interrupt firing is very straightforward after we have generated a sequence of interrupts for a symbolic execution path.
It is worth noting that actual interrupt firing is made before the interrupt firing site.

\section{Dynamic Symbolic Execution}
\label{sec:dse}
We have presented an interrupt modeling mechanism that is capable of supporting interrupts effectively for dynamic firmware analysis.
In this section, we integrate interrupt modeling with dynamic symbolic execution and demonstrate the overall infrastructure for analyzing MCU firmware.

We test MCU firmware by dynamic symbolic execution (DSE) instead of fuzzing because DSE allows us to cover complex paths that are hardly reachable by using fuzzers, such as those involving some magic number, checksum validations, and so forth.
We conduct dynamic symbolic execution on firmware binary so that we do not need the source code of MCU firmware, which is rarely available. 
In addition, presenting a solution that is completely built on top of  emulators and not relying on any real hardware also brings many benefits for us.
Prior works~\cite{zaddach2014avatar,corteggiani2018inception} conduct DSE for MCU firmware in the emulator, however, they rely on the real hardware to handle unemulated peripherals.
Since the real hardware entails setting up manually, these solutions are unscalable and hard to be fully-automated.
As for the existing approach~\cite{davidson2013fie} that can get rid of the real hardware by building simple models of peripherals, it is inefficient due to the state explosion problem even though it can be highly-automated.
Our approach, which is not tied to the real hardware, allows us to conduct DSE fully within the emulator, thereby making firmware analysis more effective, scalable, and fully-automated without suffering from the state explosion problem.

We conduct dynamic symbolic execution in angr~\cite{shoshitaishvili2016sok}. 
Angr provides an emulated environment that does not emulate any of the diverse peripherals used by the MCU firmware.
Therefore, to execute and analyze an MCU firmware, we extend the emulator to support the interrupt, as well as the memory-mapped peripheral registers, that are frequently used by the firmware. 
Specifically, we handle peripheral registers by the category-based register modeling mechanism proposed in P$^2$IM~\cite{feng2020p2im}. 
We support interrupts on demand using our interrupt modeling mechanism. 
Besides that, we also add a path scheduling algorithm to angr in order to improve the efficiency of performing dynamic symbolic execution and it will be discussed in detail later.

The dynamic symbolic execution process is interleaved with the interrupt modeling process, which consists of two components: interrupt identification and just-in-time interrupt sequence inference and interrupt firing.
We start the symbolic execution from the entry point of an MCU firmware.
During dynamic symbolic execution, when an interrupt line is enabled, we suspend symbolic execution and switch to the interrupt identification procedure to identify the enabled events multiplexed on the interrupt line and their effects on global variables. 
We resume the symbolic execution once the interrupt identification has been finished. 
We switch to the interrupt identification procedure again when the configuration of events for a peripheral is changed and there are new events being enabled.
When the firmware reads a global variable that can be modified due to interrupt firing, we suspend dynamic symbolic and switch to just-in-time interrupt sequence inference and interrupt firing. 
After firing interrupts, we resume the dynamic symbolic execution for an MCU firmware.

Since the raw input of a firmware is taken from the data registers (\texttt{DR}) of peripherals, we decide to symbolize the values of \texttt{DR} so as to conduct dynamic symbolic execution. 
MCU firmware also takes input from the interrupt, hence we symbolize the global variable to conduct a local-scope symbolic execution for about 30 basic blocks to infer interrupt sequences (explained in \S\ref{sec:int_firing}).
In this way, we are able to cover diverse paths due to the different values of global variables.
However, once we have inferred an interrupt sequence and fire interrupts accordingly, we terminate the local-scope symbolic execution and make the value of the global variable become a concrete value. 
Therefore, the symbolic execution that aims to infer interrupt sequences is different from the global symbolic execution that symbolizes the data registers since the latter leverages symbolic values throughout the whole process of symbolic execution. 

To detect memory errors triggered during DSE, we implement an error checker that enforces the least permissions needed by the firmware to memory regions.
Specifically, the checker grants read+execute permission for flash region, read+write permission for ram and peripheral region, and no permission for the rest of memory space.
Any access by the firmware that violates any of these permissions will be trapped by the checker and flagged as an error.

A common problem of dynamic symbolic execution is the state explosion problem, where the number of paths grows exponentially during symbolic execution and becomes intractable. 
In our dynamic symbolic execution, not only can the symbolic values from \texttt{DR} fork the execution into different paths, but also the interrupt sequence inference and firing can generate extra interrupt sequences or paths. This will subsequently lead the symbolic execution to keep forking more and more paths. 
To tackle this problem, we design a path prioritization mechanism based on code coverage metrics. 
We assign higher execution priority to paths that are more likely to cover new basic blocks, which are not covered previously. 
Specifically, for each path, we traverse the control flow graph (CFG) from the path's current position to calculate its minimal distance to a basic block that has never been covered before. 
The distance is denoted by how many jumps or branches (i.e., edges on CFG) are needed to reach the basic block. 
We leverage this denotation to prioritize the testing of diverse paths thereby improving the efficiency of firmware testing.
\section{Implementation}
We implement our dynamic firmware analysis framework AIM on \texttt{angr} binary analysis platform version 8.20.7.27~\cite{shoshitaishvili2016sok}. 
\texttt{Angr} conducts dynamic symbolic execution (DSE) in an emulated environment using a hybrid of symbolic and concrete values for memory and CPU registers.
We use the default SimEngine of \texttt{angr} (and leave the usage of unicorn engine, which can potentially improve the analysis performance, as future work). 
As \texttt{angr} supports the instruction set of ARM Cortex-M MCUs (i.e., ARM Thumb-2) but not the peripherals, we extend \texttt{angr} to support the interrupt, apart from other peripheral interfaces (e.g., memory-mapped registers), to enable DSE on MCU firmware binary.

We implement our framework by the flexible python interface provided by \texttt{angr}.
Our implementation consists of 2,717 lines of python code (counted by cloc~\cite{cloc} with comment lines excluded), among which 2,012 lines are for the interrupt modeling module and 705 lines are for performing DSE.
For interrupt modeling, we implement the interrupt identification and interrupt inference and firing components by 1,160 and 611 lines of code respectively.

It is worthwhile to mention that we need to emulate NVIC, the interrupt controller that is responsible for managing interrupt enabling, disabling, and firing on ARM Cortex-M MCU, within an emulator to facilitate interrupt modeling.
NVIC emulation is virtually required by any emulator-based MCU testing mechanisms that need to handle interrupts~\cite{davidson2013fie,feng2020p2im,zhou2021automatic,scharnowskifuzzware}.
We emulate NVIC by referring to the MCU architecture technical reference manual~\cite{cortexm3-nvic} and it costs 241 lines of code.
However, NVIC emulation is only a one-time effort since NVIC is generic across all ARM Cortex-M MCUs. 
We also add other miscellaneous support to \texttt{angr}, such as VTOR register of system control block that stores the base address of the interrupt vector table. 

Although \texttt{angr} supports basic DSE, we still have to extend it for testing MCU firmware and conquering some limitations. 
First, as we have known that \texttt{angr} does not emulate any MCU peripherals, which makes a firmware unable to execute, it entails us adding the support of both peripheral interfaces, namely interrupts and memory-mapped registers.
We handle interrupts via our proposed interrupt modeling mechanism.
To handle registers, we port the register modeling mechanism proposed in P\textsuperscript{2}IM~\cite{feng2020p2im} to \texttt{angr} for running DSE. 
This costs 198 lines of code in total.
It is worth noting that, due to the modular design of our framework, we can utilize other register modeling mechanisms, such as $\mu EMU$~\cite{zhou2021automatic}, to handle registers in lieu of P\textsuperscript{2}IM. 
Second, we implement a memory error detector, as illustrated in \S\ref{sec:dse}, by 52 lines of code to capture memory errors triggered during DSE.
Specifically, we intercept each memory access using memory read and write breakpoints provided by \texttt{angr}. 

Apart from the essential components for testing MCU firmware, we also need to improve the efficiency of performing DSE by solving the state explosion problem.
\texttt{Angr} utilizes a trivial breadth-first search mechanism for path scheduling, which gives all paths the same execution priority.
This strategy will make the execution of DSE quickly fall into state explosion, thereby hindering us from testing MCU firmware adequately. 
As a result, we implement a coverage-based path scheduling mechanism (explained in \S\ref{sec:dse}) utilizing the control-flow graph generated by \texttt{angr}'s built-in CFG building algorithm. 
The path scheduling consumes 121 lines of code. 
The rest lines of code boil down to utility and miscellaneous functionalities.
\section{Evaluation}
\label{sec:eval}
We evaluated our dynamic firmware analysis framework from two aspects: 
the effectiveness of our interrupt modeling mechanism and dynamic symbolic execution results. 

\subsection{Methodology}
We run our experiments against a set of real MCU firmware for a period of seven days.
The experiments are conducted on a server equipped with a 10-core Intel Xeon Silver 4114 CPU@2.20GHz, 276 GB of RAM, and Ubuntu 16.04. 

\vspace{0.2em}
\noindent\textbf{Test Suite:}
We utilize the firmware dataset used in P\textsuperscript{2}IM~\cite{feng2020p2im} to evaluate our framework.
The dataset is composed of ten full-fledged firmware that contain all common components, such as kernel, drivers, libraries, application logic, and console.
We are able to test eight of them in total, namely CNC, Gateway, Heat Press, PLC, Reflow Oven, Steering Control, Drone, and Robot.
Their functionalities are very diverse, ranging from industrial control systems to gateways. 
We remove two firmware (i.e., Soldering Iron and Console) since they are built upon real-time operating systems, which use the instructions not supported by \texttt{angr}.

\vspace{0.2em}
\noindent\textbf{Baseline:}
To measure how much improvement our framework can achieve specifically, we set up two baseline implementations, namely Baseline-No\_INT and Baseline-P\textsuperscript{2}IM, to perform DSE. 
They represent the bare-minimum and the state-of-the-art implementations of emulator-based firmware analysis framework respectively.
Specifically, Baseline-No\_INT does not provide any interrupts while conducting DSE, whereas Baseline-P\textsuperscript{2}IM adopts a naive interrupt modeling approach that is used by state-of-the-art emulator-based firmware testing mechanisms including P$^2$IM~\cite{feng2020p2im}, Laelaps~\cite{cao2020device}, $\mu EMU$\cite{zhou2021automatic}, Jetset\cite{johnson2021jetset}, and Fuzzware\cite{scharnowskifuzzware}. 
The naive interrupt model fires interrupts at a fixed order and frequency (i.e., every 1,000 basic blocks). 
To build Baseline-P\textsuperscript{2}IM, we extend  P\textsuperscript{2}IM~\cite{feng2020p2im} so that it can analyze firmware not only by fuzzing but also by DSE.

\subsection{Interrupt Modeling Statistics}
We demonstrate the effectiveness of our proposed interrupt modeling mechanism according to interrupt identification and interrupt sequence inference.

\subsubsection{Interrupt Identification}

\begin{table*}[!ht]
    \centering
    \caption{The statistics of interrupt identification}
    \label{tab:isr_analysis_stats_long}
    \begin{tabular}{|l|c|c|c|c|c|c|c|wc{1.4cm}|c|c|c|}
    \hline
    \multirow{3}{*}{\textbf{Firmware}}  & \multicolumn{2}{c|}{\multirow{2}{*}{\textbf{Interrupt Lines}}} & \multicolumn{3}{c|}{\multirow{2}{*}{\textbf{Events}}} & \multicolumn{2}{c|}{\textbf{Before Optimization}} & \multicolumn{4}{c|}{\textbf{After Optimization}} \\
    \cline{7-8}\cline{9-12}
     ~ & \multicolumn{2}{c|}{} & \multicolumn{3}{c|}{} & \multirow{2}{*}{\textbf{\# Paths}} & \multirow{2}{*}{\textbf{\makecell{\# Global\\ Variables}}} & \multirow{2}{*}{\textbf{\makecell{\# Filtered \\ Paths}}} & \multicolumn{3}{c|}{\textbf{Global Var. Path \#}} \\
     \cline{2-3}\cline{4-6}\cline{10-12}
        ~ & Total & Used & Total & Enabled & per Line & ~ & ~ & ~ &  Min. & Max. & Avg. \\ 
        \hline
        \textbf{PLC} & 91 & 2 & 8 & 7 & 3.5 & 156 & 10 & 12 & 1 & 10 & 2.2 \\ \hline
        \textbf{Gateway} & 43 & 4 & 20 & 8 & 2 & 203  & 12 & 15 & 1 & 10 & 2.1 \\ \hline
        \textbf{Reflow O.} & 43 & 2 & 8 & 8 & 4 & 155 & 12 & 13 & 1 & 10 & 2.1 \\ \hline
        \textbf{Heat Press} & 45 & 2 & 3 & 3 & 1.5 & 5 & 3 & 3 & 1 & 1 & 1 \\ \hline
        \textbf{Steering C.} & 45 & 3 & 4 & 4 & 1.3 & 6  & 4 & 4 & 1 & 1 & 1 \\ \hline
        \textbf{CNC} & 91 & 2 & 4 & 1 & 0.5 & 9  & 1 & 2 & 1 & 1 & 1 \\
        \hline
        \textbf{Drone} & 43 & 2 & 8 & 6 & 3 & 36 & 11 & 19 & 1 & 17 & 2.8 \\
        \hline
        \textbf{Robot} & 43 & 1 & 1 & 1 & 1 & 1 & 1 & 1 & 1 & 1 & 1 \\
    \hline
    \end{tabular}
\end{table*}

The first component of our interrupt modeling aims to identify the enabled interrupt lines and events of a tested MCU firmware, and characterize the effects of interrupts on the firmware. 
In this part, we show the interrupt identification results of eight tested MCU firmware, which demonstrate why it is so challenging to provide accurate and appropriate interrupts for DSE.

Table~\ref{tab:isr_analysis_stats_long} shows statistics of interrupt identification for testing MCU firmware for seven days. 
First, even though all MCU firmware include many interrupt lines, they only utilize a very small portion of them. 
For example, Gateway, which utilizes the most percentage of interrupt lines, only enables 4/43 (or 9.3\%) of interrupt lines.
It makes use of interrupts to report time ticking (\texttt{SysTick} interrupt), make serial port communications (\texttt{USART} interrupt), and do bus transactions (\texttt{I2C} interrupt).
As for events, MCU firmware, except for Gateway, Drone and CNC, enables nearly all events multiplexing on active interrupt lines.
Also, the average number of enabled events multiplexing on an interrupt line varies from less than 1 to 4.
For example, a \texttt{USART} interrupt line can multiplex events for data receiving, data transmission, and error reporting. 
In short, MCU firmware can dynamically enable or disable interrupt lines and their events.
To trigger an interrupt, one has to carefully choose both the interrupt line and a specific event.
Only enabled interrupt lines and events can be used to trigger interrupts,  otherwise the firmware will crash.
Hence, interrupt triggering itself is complex making the support of interrupts a non-trivial thing.
In consequence, none of the existing works can support interrupts in such an accurate, fine-grained way as what we have accomplished. 

Next, our framework analyzes a firmware's ISRs to collect the unique event handling paths for all possible interrupts. 
Through further analysis, our framework can also obtain a systematic understanding regarding the effects of invoking those paths (e.g., modifications toward global variables).
As shown in Table~\ref{tab:isr_analysis_stats_long}, we list the number of ISR paths and the corresponding global variables for each tested MCU firmware before and after path optimization.
It is clear that PLC, Gateway, and Reflow Oven have more complicated event handling logic.
Specifically, we can generate more than 155 unique ISR paths for the enabled interrupt lines and events initially. 
ISR paths of each firmware collectively can modify more than 10 global variables.
It is noteworthy that the large number of ISR paths is due to the fact that ISR sequentially checks whether the interrupt is triggered by each of the events multiplexed on the interrupt line.  

To avoid wasting time on duplicated ISR paths, we cut down the number of ISR paths by removing those that cause the same effects on global variables.
This optimization allows us to focus on at most 19 ISR paths among all tested firmware.
Therefore, our framework can effectively cover an extensive set of ISR paths used to handle all possible interrupts.
By path optimization, we reduce the number of potential paths to search  by as many as 13 times, making the task of interrupt sequence inference much easier.

Finally, after optimization, each global variable is modified by 1 to 17 paths, which indicates the number of candidate ISR paths to select from while determining the sequence of interrupts. The rationale is that we try to choose an ISR path that might cause minimal side effects. 
For each firmware, a global variable may be modified by up to 2.8 paths on average. 
As a result, it is much more difficult to support interrupts for PLC, Gateway, Reflow Oven, and Drone than the others.

\subsubsection{Interrupt Sequence Inference \& Firing}

\begin{table}[!t]
    \centering
    \caption{The statistics of interrupt sequence inference and firing}
    \label{tab:int_firing_stats}
    \resizebox{\columnwidth}{!}{%
    \begin{tabular}{lcccrccrr}
    \toprule
      \multirow{2}{*}{\textbf{Firmware}} &
      \multirow{2}{*}{\textbf{\makecell{Int. Fir. \\ Sites}}} &
      \multicolumn{3}{c}{\textbf{Int. Sequence}} &
      \multicolumn{3}{c}{\textbf{Optimized Int. Seq.}} & \multirow{2}{*}{\textbf{Impr.}}\\
        ~ & ~ & Min. & Max. & Avg. & Min. & Max. & Avg. & ~ \\ 
        \midrule
        \textbf{PLC} & 2,635 & 1 & 9 & 3.8 & 1 & 2 & 1.4 & 2.7x \\
        \textbf{Gateway} & 2,467 & 1 & 43 & 7 & 1 & 11 & 1 & 7x \\
        \textbf{Reflow O.} & 9,073 & 1 & 3 & 1.7 & 1 & 3 & 1 & 1.7x \\ 
        \textbf{Heat Press} & 1,422 & 1 & 8 & 2.5 & 1 & 3 & 1.2 & 2.1x \\ 
        \textbf{Steering C.} & 14,079 & 1 & 3 & 1.7 & 1 & 2 & 1 & 1.7x \\ 
        \textbf{CNC} & 5 & 1 & 1 & 1 & 1 & 1 & 1 & - \\
        \textbf{Drone} & 434 & 1 & 3 & 1.1 & 1 & 1 & 1 & 1.1x \\
        \textbf{Robot} & 9 & 1 & 1 & 1 & 1 & 1 & 1 & 1x \\
        \hline
        \textbf{Average} & 3,766 & ~ & ~ & 2.3 & ~ & ~ & 1.07 & 2.2x \\ 
        \bottomrule
    \end{tabular}%
}
\end{table}

We show the statistics of interrupt inference and firing for eight tested firmware in Table~\ref{tab:int_firing_stats}.
First, the tested firmware vary greatly in that the number of places a firmware asks for any interrupt can range from five for CNC to 14,079 for Steering Control.
In other words, CNC and Steering Control represent the firmware that rely on the interrupt the least and most frequently.
However, the firmware that has more complex event handling logic, such as PLC, Gateway, and Reflow Oven, is not necessarily to be more dependent on the interrupt.
On the contrary, Steering Control, which has only 4 unique ISR paths for handling interrupts, needs interrupts at more than 14K distinct places within its firmware.
To the best of our knowledge, our framework is the first work that can accurately identify interrupt firing sites before supporting any interrupts.

Second, for each interrupt firing site, we generate multiple different interrupt sequences each of which can drive an MCU firmware toward a unique execution path.  
In Table~\ref{tab:int_firing_stats}, we show the minimum, maximum and average number of interrupt sequences we have inferred for all tested firmware among different interrupt firing sites.
We have inferred a total of 70,598 interrupt sequences (i.e., about 2.3 per site).
In a simple case, all MCU firmware contain an interrupt firing site where its subsequent execution follows only one path regardless of whether and what interrupts are supported.
For the most complicated case, we generate as many as 43 interrupt sequences at one site for Gateway.
In particular, Gateway, which handles interrupts complicatedly, enables the most number of interrupt lines and events, and it also contains the most complicated event handling logic due to using 203 ISR paths. 
Here, we notice that a single interrupt firing site for Gateway can lead to seven different execution paths on average.
In short, compared to existing works, our framework is capable of testing MCU firmware code at deeper places, which rely on more complex interrupt sequences to reach. 
This is hard to achieve by even the state-of-the-art approach like \cite{feng2020p2im,cao2020device,zhou2021automatic,johnson2021jetset,scharnowskifuzzware} since they merely fire interrupts at a fixed order and frequency.

Although we have made our interrupt inference efficient in that we only generate one interrupt sequence per firmware execution path, we still could encounter the state explosion problem.
That is to say, as we test MCU firmware deeper, the number of paths grows exponentially causing DSE hard to proceed.
Therefore, we have added a path filtering mechanism to remove paths that test previously covered basic blocks. 
In Table~\ref{tab:int_firing_stats}, we show the statistics for the number of generated interrupt sequences (or paths) among all interrupt firing sites after path filtering.
Specifically, our framework only needs to fire about 1.07 interrupt sequences per interrupt firing site to cover the basic blocks that are previously reachable after executing multiple paths. 
In the best case, we can reduce the number of interrupt sequences by about 7 times.
The results demonstrate that we mitigate the state explosion problem remarkably without sacrificing testing performance (e.g., basic block coverage).

\subsubsection{Distribution of Testing Time}
In our experiments, we execute and test each firmware for seven whole days (168 hours). 
Therefore, we would like to know how the time is distributed.
Table~\ref{tab:overhead} shows the time spent in interrupt modeling and dynamic symbolic execution for all tested firmware. 
First, the majority of firmware analysis time is occupied by either interrupt inference \& firing or DSE, whereas the time overhead for interrupt identification is negligible, which is 0.13\% on average.
About a half (44.6\%) of firmware analysis time is used for interrupt inference \& firing, which is an essential part of our framework.
It means that although we have devised an effective interrupt modeling mechanism, the support of interrupts still costs lots of firmware analysis time.
Second, according to Table~\ref{tab:isr_analysis_stats_long}, PLC, Gateway, and Reflow Oven leverage interrupts more extensively than the rest of the tested firmware.
It turns out that nearly all the testing time of CNC and Robot, which make the least use of the interrupt, can be allocated to dynamic firmware analysis (i.e., symbolic execution).
However, for the tested firmware other than CNC, the time overhead of interrupt inference \& firing and DSE seem to be uncorrelated to how much they rely on the interrupt.
For example, it costs from 56\% to 97.6\% of testing time for firmware that heavily rely on the interrupt (e.g., PLC, Gateway, and Reflow Oven) to conduct interrupt inference \& firing.
In summary, although the interrupt inference \& firing seems time-consuming, it empowers our mechanism to cover much more code that is otherwise impossible to be covered without advanced interrupt modeling mechanism (we will demonstrate it in \S\ref{sec:dse_result}).

\begin{table}[!t]
    \centering
    \caption{The time distribution in firmware analysis}
    \label{tab:overhead}
    \resizebox{\columnwidth}{!}{%
    \begin{tabular}{lrrrrrr}
    \toprule
        \multirow{2}{*}{\textbf{Firmware}} & \multicolumn{2}{c}{\textbf{Int. Ident.}} & \multicolumn{2}{c}{\textbf{Int. Infer. \& Fir.}} & \multicolumn{2}{c}{\textbf{Symbolic Exec.}} \\ 
        ~ & Time (h) & \% & Time (h) & \% & Time (h) & \% \\ \midrule
        \textbf{PLC} & 0.03 & 0.02 & 164.0 & 97.6 & 4.0 & 2.4 \\ 
        \textbf{Gateway} & 0.25 & 0.15 & 94.0 & 56.0 & 73.8 & 43.9 \\ 
        \textbf{Reflow O.} & 0.30 & 0.18 & 127.1 & 75.7 & 40.6 & 24.2 \\ 
        \textbf{Heat Press} & $\approx$ 0 & $\approx$ 0 & 159.4 & 94.9 & 8.6 & 5.1 \\ 
        \textbf{Steering C.} & 0.01 & $\approx$ 0 & 55.0 & 32.7 & 113.0 & 67.3 \\ 
        \textbf{CNC} & 0.01 & 0.01 & $\approx$ 0 & $\approx$ 0 & 168.0 & 100.0 \\
        \textbf{Drone} & 0.01 & 0.01 & 0.24 & 0.1 & 167.8 & 99.9 \\
        \textbf{Robot} & 1.12 & 0.67 & $\approx$ 0 & $\approx$ 0 & 166.9 & 99.3 \\
        \hline
        \textbf{Average} & 0.22 & 0.13 & 75.0 & 44.6 & 92.8 & 55.3 \\
        \bottomrule
    \end{tabular}
    }
\end{table}

\subsection{Firmware Testing Performance}
\label{sec:dse_result}
In this section, we present the performance of firmware testing that is conducted through dynamic symbolic execution.
We first compare the performance of our work with two baselines and then analyze the testing trend over time.

\begin{table*}[!ht]
    \centering
    \caption{Basic block coverage of \sysnameaim with comparison to baselines}
    \label{tab:coverage}
    \begin{tabular}{lcrrrrrrcrrrrr}
    \toprule
        \multirow{2}{*}{\textbf{Firmware}} & \textbf{Total BBL} & \multicolumn{2}{c}{\textbf{Baseline-N}} & \multicolumn{2}{c}{\textbf{Baseline-P*}} & \multicolumn{2}{c}{\textbf{AIM}} & \multirow{2}{*}{} & \multicolumn{2}{c}{\textbf{$\Delta(P-N)$}} & \multicolumn{2}{c}{\textbf{$\Delta(A-N)$}} & \multirow{2}{*}{\textbf{Improv.}}\\ 
        ~ & \# & \# & \% & \# & \% & \# & \% & ~ & \# & \% & \# & \% & ~ \\
        \midrule
        \textbf{PLC} & 1,997 & 282 & 14.1 & 466 & 23.3 & 394 & 19.7 & & 184 & 9.2 & 112 & 5.6 & $<$0x \\ 
        \textbf{Gateway} & 4,156 & 434 & 10.4 & 506 & 12.2 & 1310 & 31.5 & & 72 & 1.7 & 876 & 21.1 & 11.2x \\ 
        \textbf{Reflow O.} & 2,025 & 228 & 11.3 & 409 & 20.2 & 909 & 44.9 & & 181 & 8.9 & 681 & 33.6 & 2.8x \\ 
        \textbf{Heat Press} & 1,499 & 212 & 14.1 & 416 & 27.8 & 416 & 27.8 & & 204 & 13.6 & 204 & 13.6 & 0x \\ 
        \textbf{Steering C.} & 2,533 & 215 & 8.5 & 339 & 13.4 & 493 & 19.5 & & 124 & 4.9 & 278 & 11 & 1.2x \\
        \textbf{CNC} & 2,706 & 813 & 30.0 & 782 & 28.9 & 827 & 30.6 & & -31 & -1.1 & 14 & 0.5 & $>$0x \\
        \textbf{Drone} & 2,273 & 273 & 12.0 & 292 & 12.8 & 497 & 21.9 &  & 19 & 0.8 & 224 & 9.9 & 10.8x \\
        \textbf{Robot} & 1,709 & 459 & 26.9 & 459 & 26.9 & 465 & 27.2 &  & 0 & 0 & 6 & 0.4 & $+\infty$ \\
        \hline
        \textbf{Average} & 3,150 & 486 & 15.4 & 612 & 19.4 & 885 & 28.1 & & 126 & 4.0 & 399 & 12.7 &  \\
        \bottomrule
        \multicolumn{14}{l}{* We re-implement P\textsuperscript{2}IM's interrupt model to support dynamic symbolic execution other than fuzzing.}
    \end{tabular}
\end{table*}

\subsubsection{Performance \& Comparison}
We discuss the performance of testing MCU firmware via DSE in terms of basic block coverage. 
Especially, we set up two baseline implementations, which represent dynamic firmware analysis frameworks without interrupt support (Baseline-No\_Int) and with state-of-the-art interrupt support (Baseline-P\textsuperscript{2}IM) respectively.
In general, our work has achieved significantly higher code coverage than the state-of-the-art method in that we cover a higher amount of complex interrupt-dependent code residing deep within the MCU firmware.

Table~\ref{tab:coverage} shows the code coverage of our framework and two baseline methods (i.e., Baseline-No\_Int and Baseline-P\textsuperscript{2}IM) for performing DSE on eight real firmware. 
Both of the two baselines have the same settings with AIM for conducting DSE except for the way of supporting interrupts.
First, we found the basic block coverage of Baseline-No\_Int, which does not support any interrupts, is between 8.5\% and 30\%. 
It indicates that firmware testing can already analyze a part of code without interrupts.  
The main reason is that most code for firmware booting, such as initialization, does not require any interrupts.
Baseline-P\textsuperscript{2}IM has achieved slightly better basic block coverage than Baseline-No\_Int, which is between 12.2\% and 28.9\%. 
We notice that Baseline-P\textsuperscript{2}IM performs even worse than Baseline-No\_Int on CNC since it relies on the interrupt just mildly.

AIM achieves much higher basic block coverage than both of baseline implementations. 
In other words, we are capable of testing as much as 45\% of basic blocks and about 29\% of basic blocks on average across all tested firmware.
Compared to baselines, all of the newly covered basic blocks are interrupt-dependent and require one or more interrupts to execute.
We show the improvement of code coverage made by Baseline-P\textsuperscript{2}IM over Baseline-No\_Int via $\Delta(P-N)$, and AIM over Baseline-No\_Int via $\Delta(A-N)$.
Among eight real firmware, AIM achieves higher basic block coverage than Baseline-No\_Int on all firmware, whereas Baseline-P\textsuperscript{2}IM improves only on six firmware.
On average, AIM makes an improvement for about 12.7\% than Baseline-No\_Int while Baseline-P\textsuperscript{2}IM improves for about 4\%.

AIM makes the most improvement on Reflow Oven and Gateway where it covers about 34\% and 21\% more basic blocks than Baseline-No\_Int respectively.
The result clearly proves that our framework is especially helpful for MCU firmware that requires complex interrupts or is affected by interrupts in very complicated ways.  
AIM makes the least improvement on Robot and the basic block coverage improvement is merely 0.4\% higher than Baseline-No\_Int.
It shows that we can still elevate the code coverage level higher even if the firmware does not require the interrupt too much.
In short, we observe inspiring results for MCU firmware that both slightly and heavily rely on interrupts.

Finally and most importantly, by comparing with the state-of-the-art like Baseline-P\textsuperscript{2}IM, AIM covers up to 11.2 times more interrupt-dependent code which can only be executed and tested after receiving certain interrupts. 
AIM covers 21.2\% more basic blocks than Baseline-No\_Int on Gateway whereas the rate for Baseline-P\textsuperscript{2}IM is only 1.7\%.
All the improvements of AIM over Baseline-P\textsuperscript{2}IM are on the interrupt-dependent code that is impossible to be covered by state-of-the-art interrupt modeling mechanisms~\cite{feng2020p2im,cao2020device,zhou2021automatic,johnson2021jetset,scharnowskifuzzware}.

\subsubsection{Trend}
\begin{figure}[t]
    \centering
    \includegraphics[width=\columnwidth]{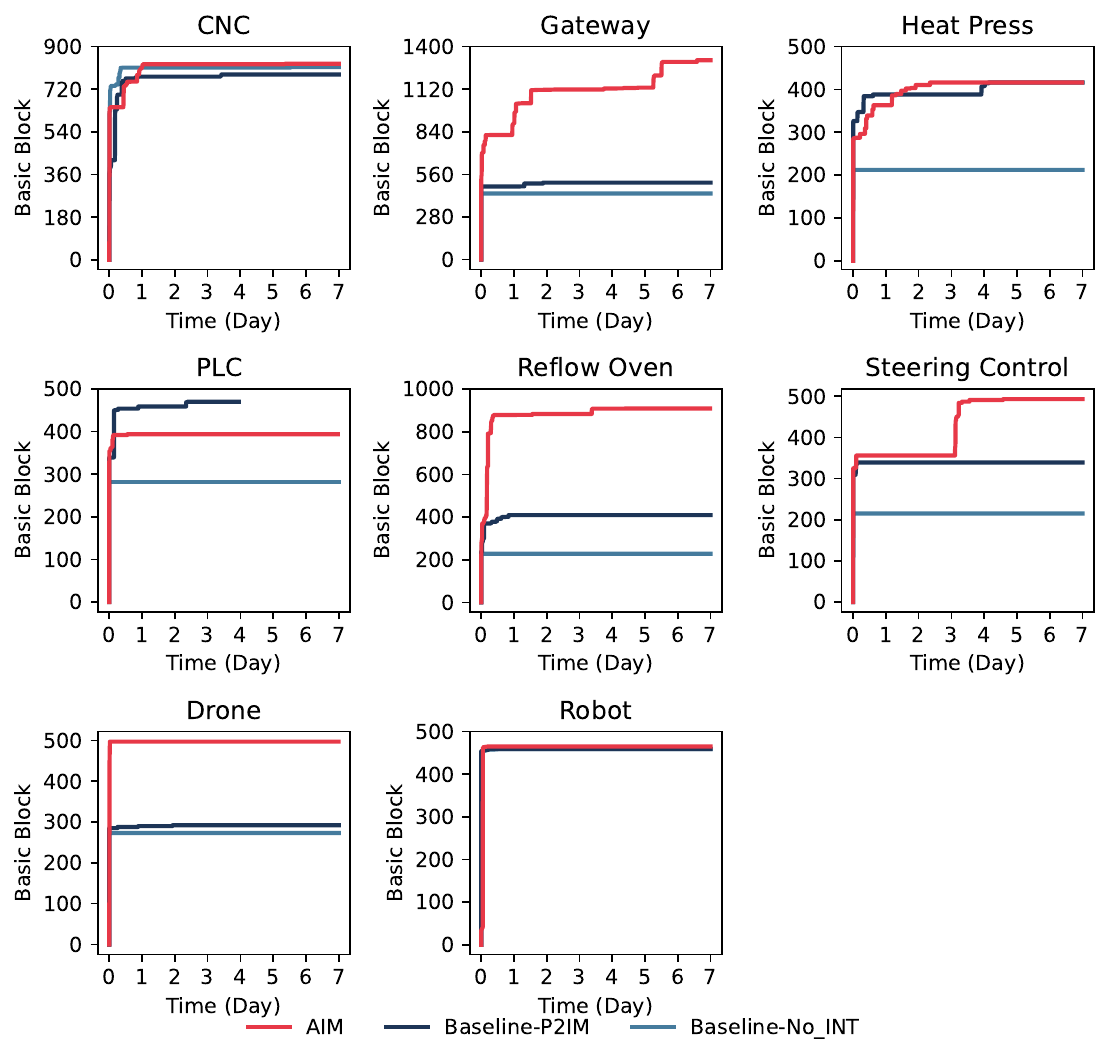}
    \caption{Basic block coverage trend over time during firmware testing}
    \label{fig:coverage_trend}
\end{figure}

In our experiments, we execute eight real firmware for a total of seven days.
During dynamic symbolic execution, the basic block coverage, path coverage, and memory consumption change over time.
As time is an important factor for DSE, understanding the performance trend is meaningful.
Figure~\ref{fig:coverage_trend} shows the general trend of basic block coverage change.
First, Baseline-No\_Int quickly reaches the best basic block coverage within a few hours and can hardly increase the coverage afterward for all firmware except for CNC.
For all firmware except for PLC, AIM keeps improving basic block coverage and finally reaches a value higher than the state-of-the-art methods.
On CNC and Heat Press, AIM starts with lower coverage than baseline-P\textsuperscript{2}IM, and surpasses it at a later time. 
It is because AIM spends more time on determining interrupt sequences and firing interrupts than baseline-P\textsuperscript{2}IM. 
For the rest of the tested firmware, our interrupt modeling mechanism is more efficient than P\textsuperscript{2}IM meaning that we can infer and fire interrupts much faster.
We observe that AIM significantly outperforms P\textsuperscript{2}IM in Gateway, Reflow Oven, Steering Control, and Drone since these firmware rely on interrupts in more complex ways.
For PLC, P\textsuperscript{2}IM exhausts all memory after about four days, causing no data to be collected afterward.

\begin{figure}[t]
    \centering
    \includegraphics[width=\columnwidth]{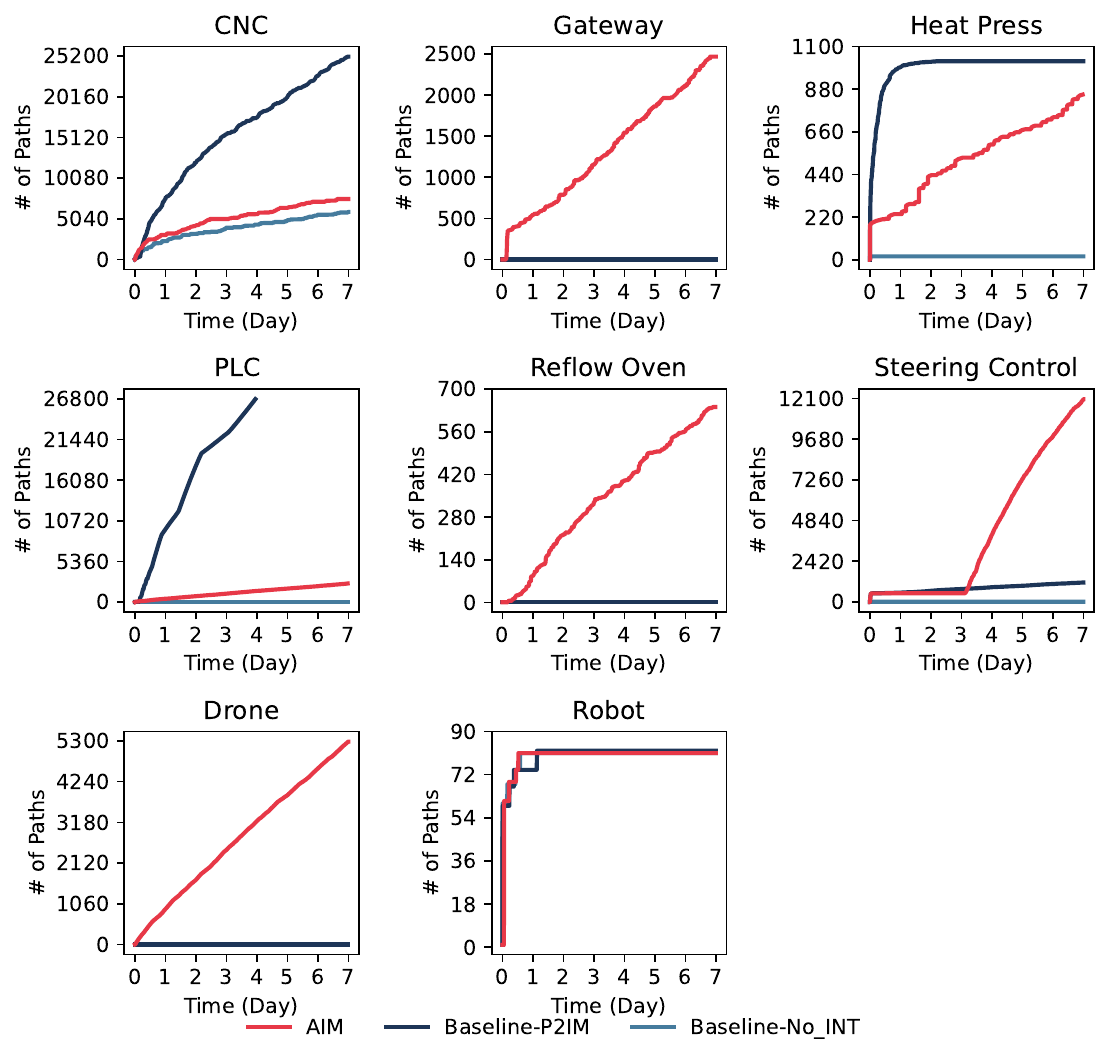}
    \caption{Path coverage trend over time during firmware testing}
    \label{fig:path}
\end{figure}

\begin{figure}[t]
    \centering
    \includegraphics[width=\columnwidth]{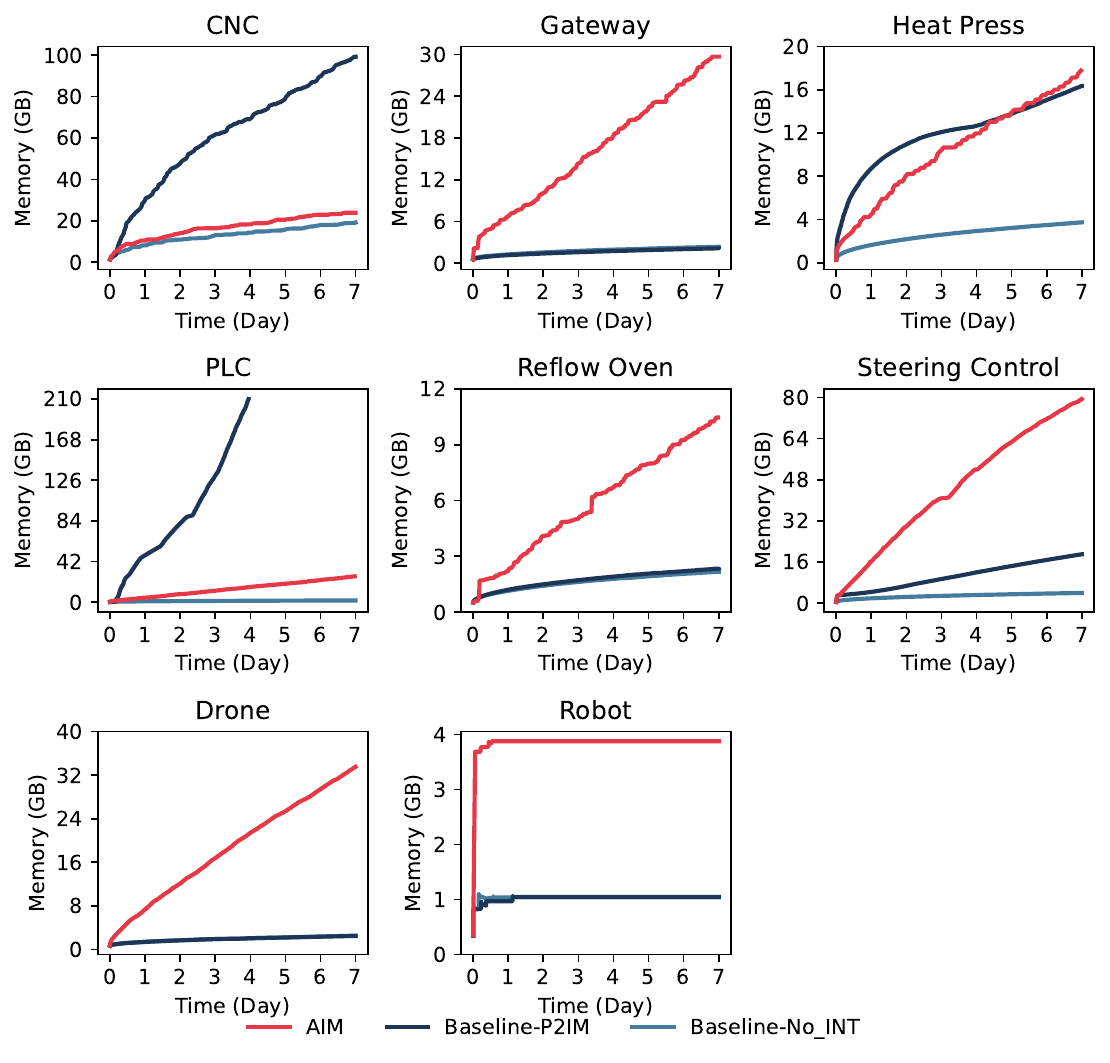}
    \caption{Memory consumption trend during firmware testing}
    \label{fig:memory}
\end{figure}

Figure~\ref{fig:path} and Figure~\ref{fig:memory} show the trend of path coverage and memory consumption during dynamic symbolic execution respectively. 
It turns out that the trend of path coverage is closely related to that of memory consumption. 
AIM consumes more memory, executes more number of paths, and covers more code than Baseline-P\textsuperscript{2}IM on Gateway, Reflow Oven, Steering Control, and Drone. 
AIM covers more code on CNC than Baseline-P\textsuperscript{2}IM, while executing less number of paths and consuming less memory. 
On Heat Press, AIM achieves the same coverage as Baseline-P\textsuperscript{2}IM while consuming comparable memory and executing less number of paths. 
On PLC, AIM falls short of the basic block coverage, memory consumption, and path coverage compared to Baseline-P\textsuperscript{2}IM.

\subsection{Comparison with State-of-the-art Firmware Fuzzer}
\label{sec:dse_vs_fuzz}
In this section, we compare the firmware testing performance of our framework with a state-of-the-art firmware fuzzer, namely Fuzzware~\cite{scharnowskifuzzware}. 
Specifically, we want to answer the following question: whether the dynamic symbolic execution-based firmware testing can uncover code that fuzzers fail to cover, or vice versa. 

Similar to AIM, Fuzzware also tests the firmware in an emulator with little peripheral support.
But there are two key differences. 
1) \textbf{Modeling of unemulated peripherals:} Fuzzware, along with other recent work~\cite{cao2020device, zhou2021automatic, johnson2021jetset}, aims to provide a more robust and precise MMIO modeling than P\textsuperscript{2}IM~\cite{feng2020p2im} to facilitate emulator-based firmware testing. 
For interrupts, it reuses the simple round-robin firing strategy from P\textsuperscript{2}IM~\cite{uEmu_interrupt_eg_round_robin} or relies on humans to specify where and what interrupts to fire~\cite{Fuzzware_interrupt_eg_manul_specify}. 
However, AIM adopts a novel just-in-time interrupt firing strategy which can automatically model the interrupts and cover more interrupt-dependent code in the firmware (\S\ref{sec:dse_result}). For MMIO, AIM simply reuses the technique from P\textsuperscript{2}IM. 
2) \textbf{Testing techniques:} Fuzzware (and other recent work~\cite{cao2020device, zhou2021automatic, johnson2021jetset}) tests firmware by fuzzing instead of dynamic symbolic execution used by AIM. Fuzzing is known to be more efficient but may fail to solve complex path constraints that dynamic symbolic execution can~\cite{stephens2016driller}.

To answer the question, we use both AIM and Fuzzware\footnote{We evaluate AIM against Fuzzware, instead of other recent works \cite{cao2020device, zhou2021automatic, johnson2021jetset}, because Fuzzware outperforms them.} to test the same eight real-world firmware from P\textsuperscript{2}IM paper (that we tested in the previous section) for 24 hours. 
Although Fuzzware is officially evaluated on this firmware suite~\cite{scharnowskifuzzware}, we are unable to reproduce the fuzzing results on a vanilla firmware binary image until we manually identify and remove some ``anti-fuzzing" functions from the firmware, such as the \texttt{delay} function in Listing~\ref{listing:delay} which causes fuzzer timeouts if a large number of timer interrupts are not fired in time.
Note that it is a common practice for existing firmware fuzzers~\cite{feng2020p2im, zhou2021automatic, scharnowskifuzzware} to remove ``anti-fuzzing" functions~\cite{Fuzzware_interrupt_eg_skip}.
However, AIM does not need to remove the ``anti-fuzzing" functions that can be easily and quickly moved forward by firing interrupts in time, such as the \texttt{delay} function. 
Indeed, AIM's interrupt modeling strategy identifies that 5 timer interrupts are needed at line 12 in Listing~\ref{listing:delay} and fires them just-in-time. 
Therefore, AIM saves the manual effort needed before launching firmware testing.

\begin{table}[]
\caption{Function coverage of AIM and the SOTA firmware fuzzer Fuzzware}
\label{tab:dse_fuzzing}
\centering

\begin{tabular}{lccccr}
\toprule
\textbf{Firmware}    & \textbf{AIM} & \textbf{Fuzzware} & \textbf{$\Delta$(A-F)} & \textbf{$\Delta$(F-A)} & \multicolumn{1}{l}{\textbf{A$\cup$F}} \\
\midrule
\textbf{PLC}         & 120          & 118               & 31                     & 29                     & 149                                   \\
\textbf{Gateway}     & 199          & 314               & 30                     & 145                    & 344                                   \\
\textbf{Reflow O.}   & 198          & 184               & 40                     & 26                     & 224                                   \\
\textbf{Heat Press}  & 103          & 91                & 30                     & 18                     & 121                                   \\
\textbf{Steering C.} & 100          & 105               & 20                     & 25                     & 125                                   \\
\textbf{CNC}         & 178          & 232               & 35                     & 89                     & 267                                   \\
\textbf{Drone}       & 113          & 174               & 11                     & 72                     & 185                                   \\
\textbf{Robot}       & 99           & 140               & 17                     & 58                     & 157                                   \\
\midrule
\textbf{Average}     & 138.8        & 169.8             & 26.8                   & 57.8                   & 196.5                   \\             
\bottomrule

\end{tabular}
\end{table}

Table~\ref{tab:dse_fuzzing} shows the evaluation results.
On average, AIM covers 138.8 functions, while Fuzzware covers 169.8 functions, 22.3\% more than AIM.
AIM achieves a higher function coverage on PLC, Reflow Oven, and Heat Press. 
Considering dynamic symbolic execution is significantly slower than fuzzing, AIM can potentially catch up if running longer than 24 hours (as shown in Figure~\ref{fig:coverage_trend}). 
AIM on average covers 26.8 unique functions that are not covered by Fuzzware (the fourth column in Table~\ref{tab:dse_fuzzing}), while Fuzzware covers 57.8 functions on average that AIM fails to cover (the fifth column in Table~\ref{tab:dse_fuzzing}). 

The last column in Table~\ref{tab:dse_fuzzing} demonstrates by combining state-of-the-art MMIO (from Fuzzware) and interrupt modeling mechanisms (from AIM), and adopting a hybrid testing approach of dynamic symbolic execution and fuzzing, we can potentially achieve more effective firmware testing with a higher overall code coverage.

\section{Related Work}

\subsection{Firmware Fuzzing}
Fuzzing, or fuzz-testing, is an automated program testing technique that is proven to be effective at finding bugs~\cite{afl,klees2018evaluating}. 
Numerous works aim to apply fuzzing to firmware by solving firmware-specific challenges. 
These works fuzz-test firmware either on a real device or in an emulator. 
When testing firmware in an emulator, the key challenge is to emulate or model the diverse peripherals available on MCUs~\cite{fasano2021sok}. 
In the rest of this section, we will organize related work by where the firmware is tested, and whether/how they handle the peripherals used by the firmware. 

\cite{chen2018iotfuzzer,li2022mu,garbelini2020sweyntooth} test firmware on a real device. 
However, fuzzing firmware on real MCU devices is severely hampered by the limited computation and storage resources of MCU. 
Therefore, recent works decide to fuzz-test firmware in emulators. 
As emulators do not provide comprehensive emulation for the diverse peripherals equipped on MCUs, these works solve this challenge in different ways. 
\cite{herdt2020verification} emulates peripherals by SystemC-based virtual prototypes, which requires significant manual effort. 

To automate the handling of unemulated peripherals in emulators, various mechanisms have been proposed. 
Prospect~\cite{kammerstetter2014prospect} and Charm~\cite{talebi2018charm} conduct hardware-in-the-loop emulation which fuzz-tests the firmware in an emulator and forwards the unemulated peripheral operations to a real device. 
Although they do not need to model interrupts (which are handled by the real device), they suffer from slow execution speed and poor scalability due to the hardware-dependency. 

A plethora of recent works fuzz-test firmware purely in an emulator by novel, automated peripheral modeling mechanisms~\cite{gustafson2019toward,spensky2021conware,feng2020p2im,mera2021dice,cao2020device,zhou2021automatic,johnson2021jetset,scharnowskifuzzware,chen2016towards,kim2020firmae,zheng2019firm,clements2020halucinator,li2021library,harrison2020partemu}. 
Pretender~\cite{gustafson2019toward} and Conware~\cite{spensky2021conware} use heuristics to infer peripheral models from the interactions they collected between the firmware and the real peripheral hardware. 
P\textsuperscript{2}IM~\cite{feng2020p2im} and DICE~\cite{mera2021dice} completely remove the need of real hardware devices (i.e., being hardware-independent) by a novel pattern-based peripheral modeling approach. 
Laelaps~\cite{cao2020device}, $\mu EMU$~\cite{zhou2021automatic}, Jetset~\cite{johnson2021jetset}, and Fuzzware~\cite{scharnowskifuzzware} model peripherals by dynamic symbolic execution-based approaches, which infer acceptable peripheral register values from the firmware. 
After generating the peripheral model, they test the firmware by fuzzing (namely, they only use dynamic symbolic execution for generating peripheral models, but not for testing the firmware like what our mechanism does). 
Among these works, \cite{feng2020p2im,mera2021dice,cao2020device,zhou2021automatic,johnson2021jetset,scharnowskifuzzware} are most closely related to our work because of sharing the same goals, which is testing generic types of firmware that uses different MCUs, peripherals, and OSes without relying on any real MCU hardware. 
However, there are two key differences. 
First, these works mainly focus on modeling peripheral registers~\cite{feng2020p2im,cao2020device,zhou2021automatic,johnson2021jetset,scharnowskifuzzware} or Direct Memory Accesses (DMA)~\cite{mera2021dice}, which are different peripheral channels than interrupts modeled by our mechanism. 
They provide very simple, coarse-grained interrupt models, which fire activated interrupts at a fixed order and frequency. 
We demonstrated in our evaluation that our interrupt outperforms these naive interrupt models. 
Second, they test firmware by fuzzing, while we conduct dynamic symbolic execution.

\cite{chen2016towards,kim2020firmae,zheng2019firm,clements2020halucinator,li2021library,harrison2020partemu} conduct hardware-independent firmware fuzzing by a different approach called high-level emulation, where they replace libraries and drivers with manually-implemented, functionality-equivalent stubs. 
Their approaches avoid firmware's accesses to the unemulated peripherals, and therefore, make it unnecessary to model such peripherals. 
Unfortunately, these approaches cannot find bugs in essential firmware components such as libraries and drivers because of the replacement. 

Unlike our mechanism which aims to generically support various types of firmware, Frankenstein~\cite{ruge2020frankenstein} and FirmWire~\cite{hernandezfirmwire} test a specific type of firmware (Bluetooth and Celluar baseband firmware respectively) by extensively using domain knowledge in emulation. 
Last, \cite{muench2018you} points out memory corruption errors triggered by fuzzers in MCU firmware can hardly cause crashes or be detected. 
\cite{salehi2020musbs} improves the detectability of memory corruptions in MCU firmware by a static binary sanitizer.

\subsection{Dynamic Symbolic Execution on Firmware}
Although dynamic symbolic execution has various benefits over fuzzing, such as solving complex path constraints or conducting more types of security analysis (e.g., reverse engineering the proprietary protocol used in the firmware), there is much less work doing dynamic symbolic execution on firmware than fuzzing. 

Avatar~\cite{zaddach2014avatar}, FIE~\cite{davidson2013fie}, and Inception~\cite{corteggiani2018inception} are dynamic symbolic execution frameworks that support generic types of MCU firmware with different functionalities. 
They handle unemulated peripherals by different approaches. 
Avatar~\cite{zaddach2014avatar} proposes a novel approach called hardware-in-the-loop emulation which tests the firmware in an emulator and uses a real hardware device to handle the unemulated peripheral operations. 
Avatar suffers from slow execution speed, poor scalability, and limited automation because of hardware dependency. 
\cite{koscher2015surrogates} and \cite{muench2018avatar} improve Avatar on the speed and flexibility of the bridge between the emulator and the real device. 
FIE~\cite{davidson2013fie} removes the hardware dependency by modeling the unemulated peripherals with simple models (which handle peripheral register reads with unconstrained symbolic value and try to fire every enabled interrupt after executing every single instruction). 
The simple peripheral model makes FIE quickly run into state explosion during dynamic symbolic execution, which severely limits the effectiveness of firmware testing. 
Inception~\cite{corteggiani2018inception} solves the state explosion problem of FIE by selectively forwarding peripheral access using a mechanism similar to Avatar, and therefore, inherits the limitations from Avatar. 

FirmUSB~\cite{hernandez2017firmusb} is a dynamic symbolic execution mechanism specifically designed for USB controller firmware (that is based on Intel 8051 MCUs). 
Since FirmUSB heavily relies on the domain knowledge of USB protocol, it cannot test firmware that is designed for different purposes or use different protocols (like our mechanism can). 

Instead of testing the whole firmware, \cite{shoshitaishvili2015firmalice} and \cite{gritti2022heapster} test specific firmware components by symbolic execution: authentication logic and heap management library (HML) respectively. 
As these components hardly have any dependency on peripherals (and their interrupts), they do not need to model interrupts like what our mechanism does. 

In summary, no existing work achieves all the goals our mechanism does: automatically supporting dynamic symbolic execution on generic MCU firmware that uses various MCU models, peripherals, and OSes. 
\section{Conclusion}
Microcontrollers or MCU are pivotal in modern IoT and embedded devices.
Similar to computer software, MCU firmware---the whole software stack of the MCU---can virtually contain diverse types of vulnerabilities. 
To run and test MCU firmware in emulators, one needs to emulate the operations of a wide range of MCU peripherals (or hardware), especially those through the interrupt. 
To effectively test MCU firmware, we propose a dynamic firmware analysis framework that can support unemulated MCU peripherals by advanced interrupt modeling and enables dynamic symbolic execution for firmware in a popular emulator. 
Our solution overcomes several open challenges, including hardware dependence and the complexity of interrupts. 
The results show our approach can cover up to 11.2 times more interrupt-dependent firmware code than state-of-the-art solutions. 
The proposed method allows us to avoid using real hardware and tests firmware logic that heavily depends on interrupts. 
Also, we achieved both remarkable performance improvement and the design goals infeasible to fulfill previously.



\ifCLASSOPTIONcompsoc
  \section*{Acknowledgments}
  This work was partially supported by National Natural Science Foundation of China grant 124009-N72402. 
\else
  \section*{Acknowledgment}
\fi


\ifCLASSOPTIONcaptionsoff
  \newpage
\fi



%
\bibliographystyle{IEEEtran}
\bibliography{bare_jrnl_compsoc}
%
\begin{IEEEbiography}[{\includegraphics[width=1in,height=1.25in,clip,keepaspectratio]{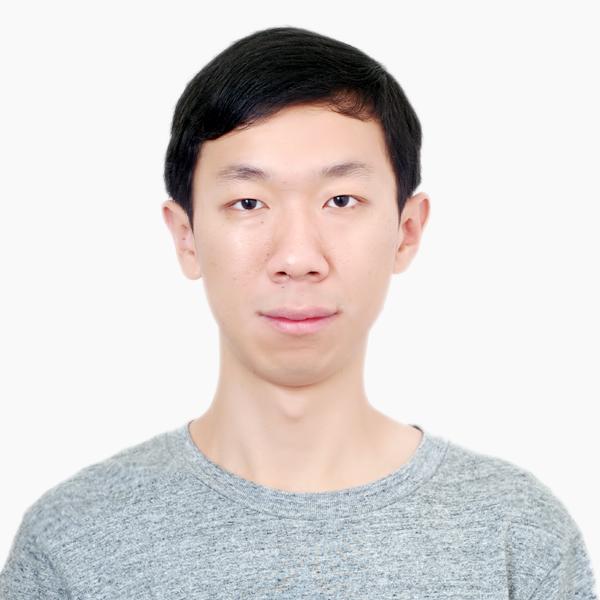}}]{Bo Feng}
is a Postdoctoral Fellow in the School of Cybersecurity and Privacy, College of Computing, Georgia Institute of Technology, USA. He received his bachelor’s degree in computer science from Wuhan University, China in 2015, and his Ph.D. degree in computer science from Khoury College of Computer Sciences, Northeastern University, USA in 2022. His research interests include system security, IoT security, and embedded devices. 
\end{IEEEbiography}

\begin{IEEEbiography}[{\includegraphics[width=1in,height=1.25in,clip,keepaspectratio]{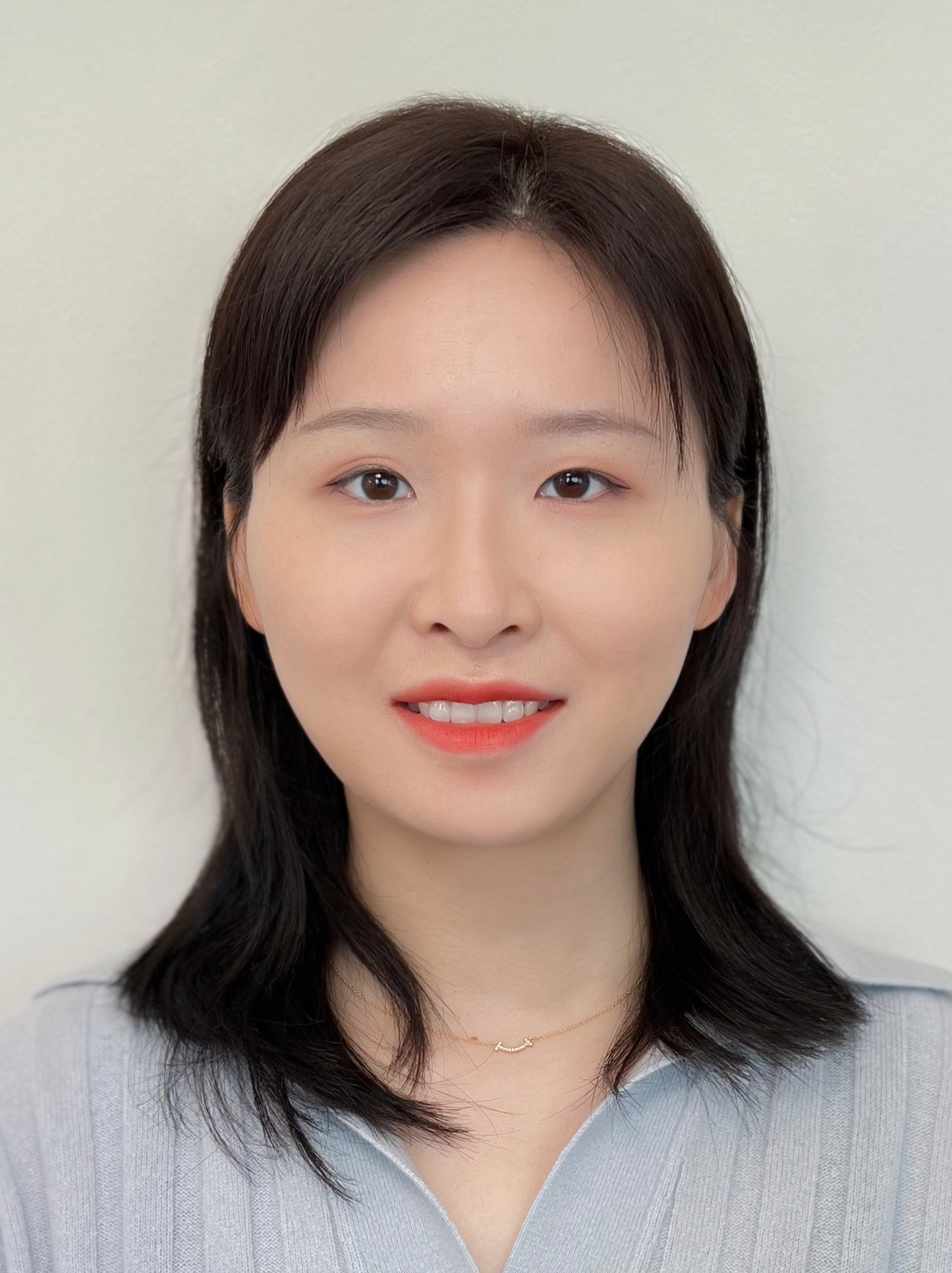}}]{Meng Luo}
is a ZJU 100-Young Professor at School of Cyber Science and Technology, Zhejiang University, Hangzhou, China since 2022. She received her B.Eng. degree in information security from Wuhan University, Wuhan, China, in 2015, and her Ph.D. degree in computer science from Stony Brook University, Stony Brook, NY, USA, in 2020. She was a Post-Doctoral Research Associate with the Khoury College of Computer Sciences, Northeastern University, Boston, MA, USA. 
Her research interests include mobile security, web security, and IoT security.
\end{IEEEbiography}

\begin{IEEEbiography}[{\includegraphics[width=1in,height=1.25in,clip,keepaspectratio]{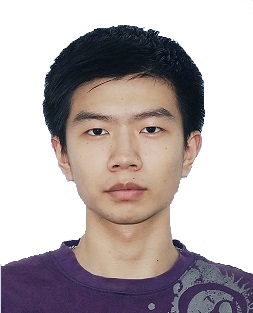}}]{Changming Liu}
is currently a PhD candidate at Northeastern University, Boston, MA. 
He is co-advised by Professor Long Lu and Engin Kirda.
Previously he got his bachelor's and master's degree from Huazhong University of Science and Technology, China. 
He has interned at IBM Watson Research Center, Microsoft Research Asia, and Hong Kong University. 
His research interests primarily reside in securing low-level system software with fuzz testing and symbolic execution.
\end{IEEEbiography}

\begin{IEEEbiography}[{\includegraphics[width=1in,height=1.25in,clip,keepaspectratio]{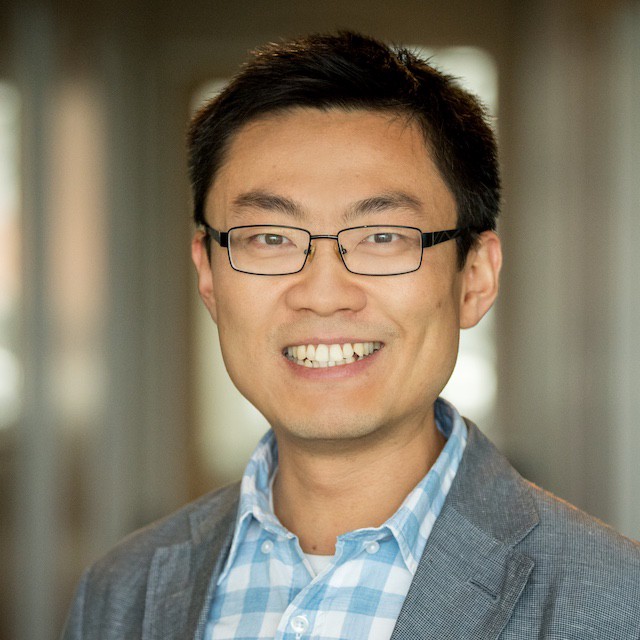}}]{Long Lu}
is an Associate Professor in the Khoury College of Computer Sciences, a core faculty member of the Cybersecurity and Privacy Institute, and a co-director for the SecLab (Systems Security Lab), at Northeastern University. His research aims to secure low-level software in widely deployed or critical systems. He designs and builds novel program analysis and hardening techniques, hardware-backed primitives for security, and trusted/confidential computing environments. His recent work has focused on embedded and IoT/CPS systems. Long has won an NSF CAREER Award, an Air Force Faculty Fellowship, two Google ASPIRE Awards, etc. His research is supported by the National Science Foundation, the Office of Naval Research, the Army Research Office, etc.
\end{IEEEbiography}

\begin{IEEEbiography}[{\includegraphics[width=1in,height=1.25in,clip,keepaspectratio]{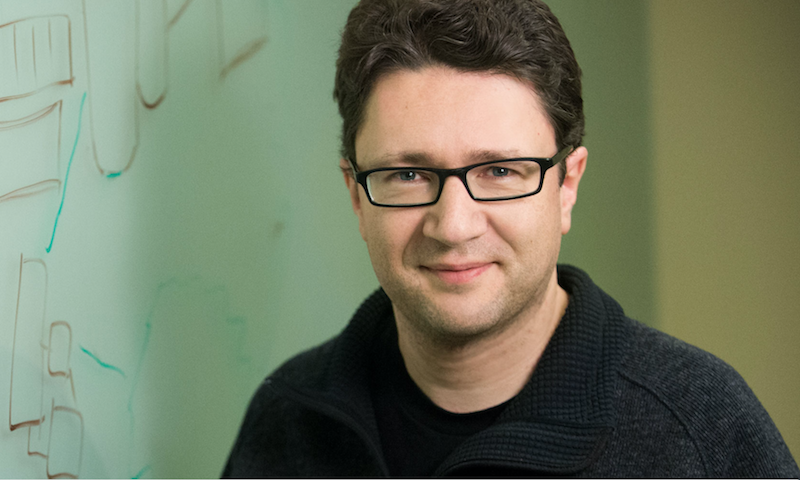}}]{Engin Kirda}
is a professor at the Khoury College of Computer Sciences and the Department of Electrical and Computer Engineering at Northeastern University in Boston. Previously, he was a tenured faculty at Institute Eurecom (Graduate School and Research Center) in the French Riviera and before that, faculty at the Technical University of Vienna where he co-founded the Secure Systems Lab. His lab has now become international and is distributed over nine institutions and geographical locations. His current research interests are in systems, software and network security (with focus on Web security, binary analysis, malware detection). Before that, he was mainly interested in distributed systems, software engineering and software architectures. He is also part of the Shellphish hacking group. They regularly participate at the DefCon CTF.
\end{IEEEbiography}








\end{document}